\documentclass[aps,prb,amsmath,twocolumn]{revtex4-2}
\usepackage{bm}
\usepackage{graphicx}

\DeclareMathOperator{\sgn}{sgn}

\DeclareMathOperator{\Rea}{\mathrm{Re}}
\DeclareMathOperator{\Sp}{\mathrm{Sp}}

\begin{document}
\title{Phase-coherent thermoelectricity and non-equilibrium Josephson current in Andreev interferometers}
\author{Mikhail S. Kalenkov}
\affiliation{I.E. Tamm Department of Theoretical Physics, P.N. Lebedev Physical Institute, 119991 Moscow, Russia}
\author{Andrei D. Zaikin}
\affiliation{Institute for Quantum Materials and Technologies, Karlsruhe Institute of Technology (KIT), 76021 Karlsruhe, Germany}
\affiliation{National Research University Higher School of Economics, 101000 Moscow, Russia}

\begin{abstract}
We develop a detailed theory describing a non-trivial interplay between non-equilibrium effects and long-range quantum coherence in superconducting hybrid nanostructures exposed to a temperature gradient.  We establish a direct relation between thermoelectric and Josephson effects in such structures and demonstrate that at temperatures exceeding the Thouless energy of our device 
both phase-coherent thermoelectric signal and the supercurrent may be strongly enhanced due to non-equilibrium low energy quasiparticles propagating across the system without any significant phase relaxation. By applying a temperature gradient one can drive the system into a well pronounced $\pi$-junction state, thereby creating novel opportunities for applications of Andreev interferometers.  

\end{abstract}
\maketitle

\section{Introduction}
It is well known that Cooper pairs can penetrate deep into a normal metal attached to a superconductor. As a result of this proximity effect, normal metals may also acquire superconducting properties \cite{deGennes,Tinkham,BWBSZ1999}.  At sufficiently low temperatures $T$ such macroscopic quantum coherence of electrons in normal metals is limited either by thermal fluctuations or by electron-electron interactions \cite{SZK,SZ15}. Accordingly, proximity induced superconducting coherence extends into a normal metal at a typical length equal 
to the shortest of two different length scales, so-called thermal length $L_T \sim \sqrt{D/T}$
(here and below $D$ is the electron diffusion coefficient) and Cooper pair dephasing length $L_\varphi$ 
that remains temperature independent \cite{SZK,SZ15,Venkat} at low enough $T$.

As a consequence of this proximity effect, macroscopic phase coherence can be established in the structure that consists
of two superconductors ($S$) connected via a normal metal ($N$) layer (with normal state resistance $R_n$) forming the so-called $SNS$ junction. 
Hence, a non-vanishing supercurrent $I_J$ may flow across such junctions which depends periodically on the superconducting phase difference $\chi$ between two superconductors. Provided the normal metal layer remains shorter that $L_\varphi$ electron-electron interactions inside it can be neglected. Then the magnitude of dc Josephson current $I_J$ in such $SNS$ junctions is controlled by the thermal length $L_T$ reaching appreciable values  $I_J \gtrsim E_{\mathrm{Th}}/(eR_n)$
in the low temperature limit \cite{ZZh,GreKa,golubov2004current} and dropping down to exponentially small values $I_J \propto e^{-\sqrt{2 \pi T/E_{\mathrm{Th}}}}$ at temperatures exceeding an effective Thouless energy $E_{\mathrm{Th}}$ of the $N$-layer.

In somewhat more complicated geometries, such as, e.g., that of an $SNS$ transistor \cite{WSZ}, one can also control both the magnitude and the sign of $I_J$ by applying an external voltage and driving the quasiparticle distribution function out of equilibrium \cite{volkov1995,WSZ,Yip,Teun}. Very recently it was demonstrated \cite{KDZ20} that yet another efficient way to control the Josephson current is to expose the junction to a temperature gradient. This situation can be realized in structures analogous to those considered in Refs. \cite{volkov1995,WSZ,Yip,Teun}, e.g., in a system composed of two superconducting and two normal terminals interconnected by normal metallic wires forming a cross. Following \cite{KDZ20} below we will denote this structure as $X$-\textit{junction}. 

Similarly to the case of voltage biased $X$-junctions \cite{volkov1995,WSZ,Yip,Teun}, by applying a temperature gradient one also drives the quasiparticle distribution function out of equilibrium. At the same there also exists an important difference between these two situations. Namely, in the voltage-biased case the Josephson current $I_J$ remains exponentially small in the high temperature limit $T \gg E_{\mathrm{Th}}$, whereas exposing an $X$-junction to a thermal gradient may yield substantial \textit{supercurrent stimulation} in this temperature range \cite{KDZ20}. Thus, applying a temperature gradient, one can effectively support long-range phase coherence in the $SNS$-type of structures at high enough temperatures where the equilibrium Josephson current already becomes negligible.

Exposing the system to a temperature gradient one also generates electric currents and/or voltages inside the sample. 
This is the essence of the so-called thermoelectric effect in superconducting structures \cite{Ginzburg}. The magnitude of this
effect becomes large as soon as electron-hole symmetry in a superconductor is violated in some way \cite{KZK12}. On top of that,
at low enough temperatures thermoelectric signals are phase-coherent, thus depending periodically on the phase of a superconducting condensate. In hybrid superconducting structures involving normal metals such phase-dependent thermoelectricity yields a variety of interesting and non-trivial effects which were extensively studied both experimentally \cite{Venkat1,Venkat2,Petrashov1,Venkat3,Petrashov2} and theoretically \cite{Sev,VH,Volk,JW,KZ17,DKZ18,dolgirev2018topology}. In addition, thermoelectric effects in superconductors give rise to a number of applications ranging from refrigeration and thermometry \cite{Giazotto} to phase-coherent caloritronics \cite{FG} and thermal logics \cite{Li} aiming to transmit information in the form of energy.

It turns out that thermoelectric and Josephson effects in the presence of a temperature gradient are intimately related to each other \cite{KDZ20}. For instance, in the case of $X$-junctions with two normal terminals kept at different temperatures $T_1$ and $T_2$ (both exceeding the Thouless energy $E_{\mathrm{Th}}$) a phase-coherent thermoelectric voltage signal $V$ generated at these terminals is related to the Josephson current  as \cite{KDZ20}
\begin{equation}
eV (\chi)  \sim eI_J(\chi )R_n \sim  E^2_{\mathrm{Th}}|1/T_1 - 1/T_2|.
\label{1}
\end{equation}
This result demonstrates that the thermoelectric voltage $V$ does not decay exponentially even in the high temperature limit  $T_{1,2}\gg E_{\mathrm{Th}}$ having exactly the same temperature dependence as $I_J$. Such non-exponential dependence of both $V$ and $I_J$ on temperature is due to the presence of non-equilibrium low energy quasiparticles suffering little dephasing while propagating through normal wires connecting two superconducting terminals.  

One can also demonstrate \cite{KDZ20} that a non-vanishing thermoelectric signal $V$ may only occur in \textit{asymmetric} $X$-junctions. The same observation holds for the non-equilibrium contribution to $I_J$ and, furthermore, a rather strong degree of the junction asymmetry is required for this contribution to reach appreciable values \cite{KZ20}. It appears, however, that the latter observations are of no general validity being merely specific to the $X$-junction geometry. 

Here we will develop a general microscopic theory of both thermoelectric and Josephson effects in superconducting hybrid structures exposed to a temperature gradient. Specifically, we will address the four-terminal hybrid structures -- frequently called Andreev interferometers -- which geometry differs from that of an $X$-junction \cite{KDZ20,KZ20}. An example of such structure is displayed in Fig. 1  

We will work out a new approach to the description of the phase-coherent transport in quasi-one-dimensional conductors interconnected between each other and attached to bulk external normal and superconducting terminals. For this purpose we reformulate the standard quasiclassical theory of superconductivity \cite{BWBSZ1999} in the spirit of Nazarov's circuit theory  \cite{Nazarov94,Nazarov99} extending the latter with the emphasis put on quasi-one-dimensional metallic conductors. 

With the aid of this approach we will set up a detailed theory describing a non-trivial interplay between proximity-induced quantum coherence and non-equilibrium effects in Andreev interferometers exposed to a temperature gradient. In particular, we will address both thermoelectric and Josephson effects which demonstrate a number of interesting new features which can be directly observed in modern experiments. We will also emphasize a close relation between these two non-equilibrium effects in the presence of a temperature gradient.

The structure of our paper is as follows. In Sec. II, we introduce our model system and describe the quasiclassical
formalism serving as a basis for our further considerations. Section III is devoted to extending the circuit theory and adopting it 
to superconducting hybrid structures under consideration. In Sec. IV, we present a detailed analysis of both thermoelectric and Josephson effects in four-terminal Andreev interferometers under the influence of a temperature gradient. In Sec. V, we briefly discuss the results and formulate our main conclusions. As usually, technical details of our calculation are relegated to the Appendixes.

\section{The model and quasiclassical equations}

Below we will consider a metallic heterostructure which consists of two superconducting and two normal terminals interconnected by five quasi-one-dimensional normal wires (of lengths $L_p$, $L_{S_{1,2}}$, $L_{N_{1,2}}$ and cross sections $\mathcal{A}_{p}$, $\mathcal{A}_{S_{1,2}}$, $\mathcal{A}_{N_{1,2}}$) as it is shown in Fig. 1. We will assume that two superconducting electrodes $S_{1}$ and $S_{2}$ are described by the order parameter values $|\Delta |\exp (i\chi_{1,2})$ and are kept at temperature $T$ and the same electrostatic potential which -- without loss of generality -- can be set equal to zero. Two normal electrodes $N_{1}$ and $N_{2}$ are, in turn, kept at different temperatures $T_{1}$ and $T_{2}$. Provided both normal terminals are disconnected from any external circuit no electric current can flow into or out of these terminals, i.e. the conditions 
\begin{equation}
I_{N_{1}}=0, \quad I_{N_{2}}=0
\label{zcc}
\end{equation}
should apply. 

In the presence of a temperature gradient quasiparticle distribution function inside normal wires is driven out of equilibrium. As a result, electrostatic potentials of two normal terminals $N_{1}$ and $N_{2}$ do not anymore equal to zero due to the thermoelectric effect. These two thermoelectric voltages -- respectively $V_1$ and $V_2$ -- induced by the temperature gradient $T_{1}-T_{2}$ will be evaluated below along with the supercurrent $I_S$ flowing between two superconducting terminals $S_{1}$ and $S_{2}$. 
\begin{figure}
\includegraphics[width=80mm]{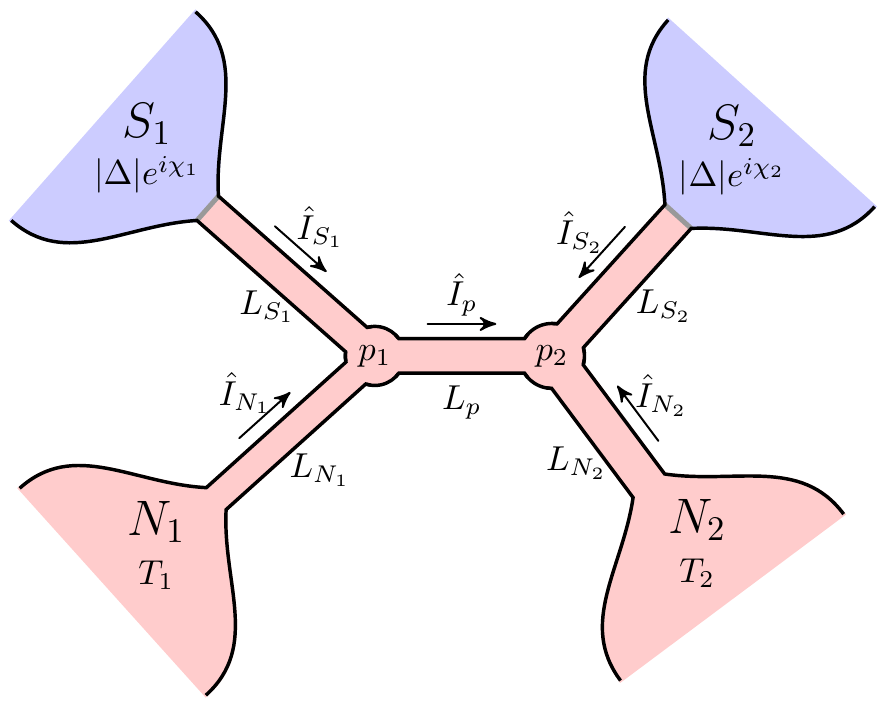}
\caption{Four terminal Andreev interferometer under consideration.}
\label{snns3-fig}
\end{figure}

Electron transport in metallic heterostructures can be conveniently described by means of quasiclassical Usadel equations \cite{BWBSZ1999} for $4\times4$ Green-Keldysh matrix functions in Keldysh$\otimes$Nambu space
\begin{equation}
\check G = 
\begin{pmatrix}
\hat G^R & \hat G^K \\
0 & \hat G^A \\
\end{pmatrix}.
\label{G}
\end{equation}
These equations read
\begin{gather}
\label{Usadel}
i D \nabla \left(\check G  \nabla \check G\right)=
\left[\check \Omega, \check G \right], \quad \check G \check G=1,
\\
\check \Omega = 
\begin{pmatrix}
\hat \Omega & 0 \\
0 & \hat \Omega \\
\end{pmatrix}, \quad
\hat \Omega=
\begin{pmatrix}
\varepsilon + eV & \Delta \\
- \Delta^* & -\varepsilon+ eV \\
\end{pmatrix}.
\nonumber
\end{gather}
Here and below $D$ is the diffusion coefficient, $\varepsilon$ labels the quasiparticle energy and $\Delta$ is the superconducting order parameter. The current density $\bm{j}$ is related to the matrix $\check G$ \eqref{G} in a standard manner as
\begin{gather}
\bm{j} =  -\dfrac{\sigma}{8 e }
\int
d \varepsilon
\Sp (\hat \tau_3\check G \nabla \check G)^K,
\end{gather}
where $\sigma$ is the normal state Drude conductivity and $\hat\tau_i$ are Pauli matrices in the Nambu space.

The Keldysh component $\hat G^K$ of the Green function matrix  \eqref{G} can be expressed via retarded ($\hat G^R$) and advanced ($\hat G^A$) components of this matrix in the form
\begin{equation}
\hat G^K = \hat G^R \hat h - \hat h \hat G^A,
\label{GKRA}
\end{equation}
where $\hat h = h^L + \hat \tau_3 h^T$ is the matrix distribution function conveniently parameterized by two different distribution functions $h^L$ and $h^T$. In normal conductors the latter functions obey diffusion-like equations
\begin{gather}
D\nabla\left[ D^T\nabla h^T + \mathcal{Y} \nabla h^L + \bm{j}_{\varepsilon} h^L \right]=0,
\label{htusadel}
\\
D\nabla\left[ D^L\nabla h^L - \mathcal{Y} \nabla h^T + \bm{j}_{\varepsilon} h^T \right]=0,
\label{hlusadel}
\end{gather}
where $D^{T,L}$ and $\mathcal{Y}$ denote dimensionless kinetic coefficients and $\bm{j}_{\varepsilon}$ represents the spectral current
\begin{gather}
D^T =\dfrac{1}{4} \Sp (1 - \hat G^R \hat \tau_3 \hat G^A \hat \tau_3)
=
\nu^2 + \dfrac{1}{4} |F^R + F^A|^2, 
\\
D^L =\dfrac{1}{4} \Sp (1 - \hat G^R \hat G^A)
=\nu^2 - \dfrac{1}{4} |F^R - F^A|^2, 
\\
\mathcal{Y} = \dfrac{1}{4} \Sp (\hat G^R \hat \tau_3 \hat G^A)
=-\dfrac{1}{4}\left(|F^R|^2 - |\tilde F^R|^2 \right), 
\\
\begin{split}
\bm{j}_{\varepsilon}=
\dfrac{1}{4}\Sp\Bigl(\hat \tau_3
\Bigl[ \hat G^R   \nabla \hat G^R - & \hat G^A \nabla \hat G^A \Bigr] \Bigr)
\\= 
\dfrac{1}{2}\Rea\Bigl( & F^R \nabla \tilde F^R - \tilde F^R \nabla F^R \Bigr),
\end{split}
\label{jE}
\end{gather}
$\nu = \Rea G^R$ is the local density of states and $G^{R,A}$, $F^{R,A}$ and  $\tilde F^{R,A}$ are components of retarded and advanced Green functions
\begin{equation}
\hat G^{R,A}
=
\begin{pmatrix}
G^{R,A} & F^{R,A} \\
\tilde F^{R,A} & - G^{R,A}
\end{pmatrix}.
\end{equation}
Note that the kinetic coefficient $\mathcal{Y}$ specifically accounts for the presence of electron-hole asymmetry in our system.
 
In the vicinity of the interfaces between normal wires and bulk metallic terminals the Green functions may change abruptly, in which case quasiclassical equations \eqref{Usadel} cannot be applied and should be supplemented by proper boundary conditions. In the case of diffusive conductors considered here these boundary conditions read \cite{Nazarov99}
\begin{multline}
\mathcal{A} \sigma_- \check G_- \nabla \check G_- =\mathcal{A}\sigma_+ \check G_+ \nabla \check G_+
\\=
\dfrac{e^2}{\pi}
\sum_{n}
\dfrac{2{\mathcal T}_n [ \check G_- , \check G_+]}{4+{\mathcal T}_n (\{ \check G_- , \check G_+\} -2)},
\label{BcNaz}
\end{multline}
where $\check G_-$ and $\check G_+$ are the Green-Keldysh matrices respectively at the left and the right sides of the interface, the sum runs over all conducting channels of the interface and ${\mathcal T}_n$ is the transmission of the $n$-th conducting channel.

In the next section we will work out a circuit theory that will allow us to establish a formal solution of the above equations for the system displayed in Fig. \ref{snns3-fig} and, more generally, for an arbitrary network of quasi-one-dimensional normal wires.

\section{Circuit theory}

Our approach is to a certain extent similar to Nazarov's circuit theory  which was originally formulated for low energy transport \cite{Nazarov94} and subsequently generalized to arbitrary energies \cite{Nazarov99}. This circuit theory, being quite general, can in principle be employed for conductors of arbitrary dimensionality. Still, in the case of spatially extended low dimensional structures it sometimes remains rather complicated for practical calculations. Below we will focus our attention on quasi-one-dimensional conductors and reformulate the quasiclassical theory of superconductivity in a form that appears more suitable both for quantitative calculations and for qualitative analysis.

\subsection{Extended conductors}
Let us rewrite Eqs. \eqref{htusadel}, \eqref{hlusadel} in the matrix form
\begin{equation}
D\nabla\left[
\hat D \nabla \hat H + \hat \tau_1 \bm{j}_{\varepsilon} \hat H
\right] = 0,
\label{diffmat}
\end{equation}
where we defined
\begin{equation}
\hat H =
\begin{pmatrix}
h^T \\
h^L \\
\end{pmatrix}, \quad
\hat D=
\begin{pmatrix}
D^T & \mathcal{Y} \\
-\mathcal{Y} & D^L\\
\end{pmatrix}.
\end{equation}
In the case of quasi-one-dimensional conductors Eq. \eqref{diffmat} yields
\begin{equation}
\hat D \hat H' +  \hat \tau_1 j_{\varepsilon} \hat H
= - e \hat I /(\mathcal{A} \sigma) .
\label{diffmat1D}
\end{equation}
where $\mathcal{A}$ is the wire cross section and 
\begin{equation}
\hat{I} =
\begin{pmatrix}
I^T \\ I^L
\end{pmatrix}
=-\dfrac{\mathcal{A} \sigma}{4 e}
\begin{pmatrix}
\Sp (\check G \check G' \hat \tau_3)^K \\ \Sp (\check G \check G')^K
\end{pmatrix}
\end{equation}
is the matrix current which remains conserved along the normal wire segment. The total electric current $I$ flowing across the wire is linked to the $I^T$-component of the matrix current by means of the following relation
\begin{gather}
I= \frac{1}{2} \int I^T  d \varepsilon .
\label{current}
\end{gather}

\begin{figure}
\includegraphics{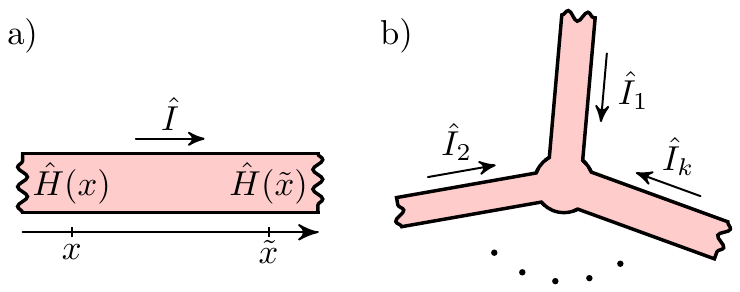}
\caption{(a) A segment of a quasi-one-dimensional normal wire and (b) several normal wires connected to each other in the node.}
\label{nwire-fig}
\end{figure}

It will be convenient for us to introduce the matrix evolution operator $ \hat U$ which obeys the equation
\begin{equation}
\hat D \hat{U}'(x,\tilde x,\varepsilon) +
\hat \tau_1 j_{\varepsilon} \hat{U}(x,\tilde x,\varepsilon)=0
\label{U}
\end{equation}
combined with the initial condition
\begin{equation}
\hat{U}( x, x,\varepsilon)=1,
\end{equation}
where $x$, $\tilde x$ are two coordinates along the normal wire. With the aid of this evolution operator we can resolve Eq. \eqref{diffmat1D} and establish the relation between the matrix current $\hat I$ and the matrix distribution function $\hat H$ at the points $x$ and $\tilde x$, cf. Fig. \ref{nwire-fig}a. It reads
\begin{equation}
\hat G_{x,\tilde x} \hat H(x) + 
\hat G_{\tilde x,x}  \hat H(\tilde x) =
- e \hat I,
\label{kirmat}
\end{equation}
where  $\hat G_{x,\tilde x}$ is $2\times 2$ conductance matrix
\begin{equation}
\hat G_{x,\tilde x} = 
\sigma j_{\varepsilon} \mathcal{A} \hat\tau_1 \left[1 - \hat U(x,\tilde x,\varepsilon)\right]^{-1}. 
\end{equation}
It obeys the following relations
\begin{equation}
\hat G_{x,\tilde x} + \hat \tau_3  \hat G_{\tilde x,x}^T \hat \tau_3 =0,
\quad
\hat G_{x,\tilde x} + \hat G_{\tilde x,x}
=
 \hat G_j,
\end{equation}
where we defined $\hat G_j = \hat\tau_1 G_j$ and $G_j = \sigma j_{\varepsilon} \mathcal{A}$.

Equation \eqref{kirmat} represents an important result. Its form closely resembles that of the standard Kirchhoff law for normal electric circuits where the matrix distribution function $\hat H$ plays the role analogous to that of the voltage. This observation enables one to operate with both matrix distribution functions $\hat H(x)$ and conductance matrices $\hat G_{x,\tilde x}$ employing the standard rules of electric engineering. For instance, in Appendix \ref{Aparallel}  we demonstrate that the conductance matrices of several elements connected either in series or in parallel can be replaced by a single equivalent conductance matrix in the same way as it is routinely done for conductances in normal electric circuits. Furthermore, in Appendix \ref{AYD} we also establish a direct matrix analogue of the standard transformation between Y-shaped and $\Delta$-shaped electric circuits.

Finally, we verify that in the absence of superconductivity (i.e. provided all terminals in our structure remain normal) the conductance matrix $\hat G_{x,\tilde x}$ reduces to a trivial form
\begin{equation}
\hat G_{x,\tilde x} = \dfrac{\sigma \mathcal{A}}{x - \tilde x} \hat 1.
\label{Rnorm}
\end{equation}

\subsection{Boundary conditions}

Let us now specify the boundary conditions at the points where the wire is connected to another wire and/or to a bulk metallic terminal.

\subsubsection{Intersection nodes}
At the nodes where several normal wires cross each other  (see Fig. \ref{nwire-fig}b) the sum of the matrix currents flowing into the node equals to zero
\begin{equation}
\sum \hat I_k =0.
\label{curcons}
\end{equation}
The same applies for the spectral currents
\begin{equation}
\sum \mathcal{A}_k j_{\varepsilon,k} =0.
\end{equation}
The matrix distribution function $\hat H(x)$ remains continuous across the node taking the same value for all wires at their end points connected to the same node.

\subsubsection{Interface barriers}
Turning now to inter-metallic interfaces we introduce the distribution functions $h_-^{T,L}$ and $h_+^{T,L}$ on both sides of the interface and rewrite Eq. \eqref{GKRA}  as
\begin{gather}
\hat G_-^K = \hat G_-^R \hat h_- - \hat h_- \hat G_-^A,
\quad
\hat h_- = h_-^L + \hat \tau_3 h_-^T,
\\
\hat G_+^K = \hat G_+^R \hat h_+ - \hat h_+ \hat G_+^A,
\quad
\hat h_+ = h_+^L + \hat \tau_3 h_+^T.
\end{gather}
With the aid of these equations we can identically transform the Keldysh component of the boundary conditions \eqref{BcNaz} to the following matrix Kirchhoff-like equation
\begin{equation}
\hat G_{-+}\hat H_-+\hat G_{+-}\hat H_+=-e \hat I,
\label{ImatrixNazarov}
\end{equation}
where 
\begin{gather}
\hat G_{+-} =
\begin{pmatrix}
g^T & g^{\mathcal{Y}} + \mathcal{A} \sigma j_{\varepsilon}/2 \\ - g^{\mathcal{Y}} + \mathcal{A}\sigma j_{\varepsilon}/2 & g^L 
\end{pmatrix},
\label{GpmNazarov}
\\
\hat G_{-+}=
-
\begin{pmatrix}
g^T & g^{\mathcal{Y}} - \mathcal{A} \sigma j_{\varepsilon}/2 \\ - g^{\mathcal{Y}} - \mathcal{A}\sigma j_{\varepsilon}/2 & g^L 
\end{pmatrix}
\label{GmpNazarov}
\end{gather}
are the interface matrix conductances,
\begin{equation}
\hat H_\pm =
\begin{pmatrix}
h_\pm^T \\
h_\pm^L \\
\end{pmatrix}
\end{equation}
define the matrix distribution functions on both sides of the barrier and 
\begin{widetext}
\begin{gather}
j_{\varepsilon}=
\dfrac{e^2}{2\pi \mathcal{A} \sigma}
\sum_n
\left[
\dfrac{{\mathcal T}_n \Sp(\hat \tau_3 [\hat G_-^R , \hat G_+^R ])}{
4 + {\mathcal T}_n ( \{\hat G_-^R , \hat G_+^R\} -2 )}
-
\dfrac{{\mathcal T}_n \Sp(\hat \tau_3 [\hat G_-^A , \hat G_+^A ])}{
4 +{\mathcal T}_n ( \{\hat G_-^A , \hat G_+^A\} -2 )}
\right],
\label{jeNazarov}
\\
g^T(\varepsilon)=
\dfrac{e^2}{2\pi}
\sum_n
\dfrac{
{\mathcal T}_n \Sp \left [
(\hat G_+^R - \hat \tau_3 \hat G_+^A \hat \tau_3)(\hat G_-^R - \hat \tau_3 \hat G_-^A \hat \tau_3)
( 4 +{\mathcal T}_n (\hat G_-^R \hat G_+^R + \hat \tau_3 \hat G_-^A \hat G_+^A \hat \tau_3 -2))
\right]
}{
\left(4 + {\mathcal T}_n ( \{\hat G_-^R , \hat G_+^R\} -2 )\right)
\left(4 + {\mathcal T}_n ( \{\hat G_-^A , \hat G_+^A\} -2 )\right)},
\label{gTNazarov}
\\
g^L(\varepsilon)=
\dfrac{e^2}{2\pi}
\sum_n
\dfrac{
{\mathcal T}_n \Sp \left [
(\hat G_+^R - \hat G_+^A)(\hat G_-^R - \hat G_-^A)
( 4 +{\mathcal T}_n (\hat G_-^R \hat G_+^R + \hat G_-^A \hat G_+^A -2 ))
\right]
}{
\left(4 + {\mathcal T}_n ( \{\hat G_-^R , \hat G_+^R\} -2 )\right)
\left(4 +{\mathcal T}_n ( \{\hat G_-^A , \hat G_+^A\} -2 )\right)
},
\label{gLNazarov}
\\
g^{\mathcal{Y}}(\varepsilon)=
-
\dfrac{e^2}{4\pi}
\sum_n
\dfrac{
{\mathcal T}_n \Sp \left[ 
8 (\hat G_-^R \hat G_+^A \hat \tau_3 + \hat G_+^R \hat G_-^A \hat \tau_3)
+
4 {\mathcal T}_n (\hat G_+^R - \hat G_-^R ) ( \hat G_+^A - \hat G_-^A )\hat \tau_3 
+
{\mathcal T}_n [\hat G_-^R , \hat G_+^R]
[\hat G_-^A , \hat G_+^A]
\hat \tau_3
\right]
}{
\left(4 + {\mathcal T}_n ( \{\hat G_-^R , \hat G_+^R\} -2 )\right)
\left(4 + {\mathcal T}_n ( \{\hat G_-^A , \hat G_+^A\} -2 )\right)
}.
\label{gMNazarov}
\end{gather}
\end{widetext}

In the tunneling limit ${\mathcal T}_n \ll 1$ Eqs. \eqref{BcNaz} reduce to Kupriyanon-Lukichev boundary conditions \cite{Kuprianov88}. In this limit the spectral supercurrent $j_{\varepsilon}$ and the interface conductances $g^{L,T,\mathcal{Y}}$ in Eqs. \eqref{ImatrixNazarov}-\eqref{GmpNazarov}  reduce to 
\begin{gather}
j_{\varepsilon}=
\dfrac{G_N}{8 \mathcal{A} \sigma}
\Sp(\hat \tau_3 [\hat G_-^R , \hat G_+^R ] - \hat \tau_3 [\hat G_-^A , \hat G_+^A ]),
\label{jeKL}
\\
g^T(\varepsilon)=
\dfrac{G_N}{8}
\Sp 
(\hat G_+^R - \hat \tau_3 \hat G_+^A \hat \tau_3)(\hat G_-^R - \hat \tau_3 \hat G_-^A \hat \tau_3)
,
\label{gTKL}
\\
g^L(\varepsilon)=
\dfrac{G_N}{8}
\Sp
(\hat G_+^R - \hat G_+^A)(\hat G_-^R - \hat G_-^A)
,
\label{gLKL}
\\
g^{\mathcal{Y}}(\varepsilon)=
-\dfrac{G_N}{8}
\Sp 
(\hat G_-^R \hat G_+^A \hat \tau_3 + \hat G_+^R \hat G_-^A \hat \tau_3).
\label{gMKL}
\end{gather}
where $G_N = (e^2/\pi) \sum_n{\mathcal T}_n $ is the normal state interface conductance.

We observe that the matrix relation between the distribution functions at the opposite sides of the interface barrier \eqref{ImatrixNazarov} has the same structure as Eq. \eqref{kirmat} we derived for a quasi-one-dimensional wire. Hence, the circuit theory developed here allows to treat both diffusive wires and interface barriers on equal footing, thereby greatly simplifying the whole consideration. With the aid of the above equations it already becomes straightforward to evaluate the quasiparticle distribution functions everywhere inside our system, as it will be demonstrated below. 

\subsubsection{Subgap electron transport}
\begin{figure}
\includegraphics{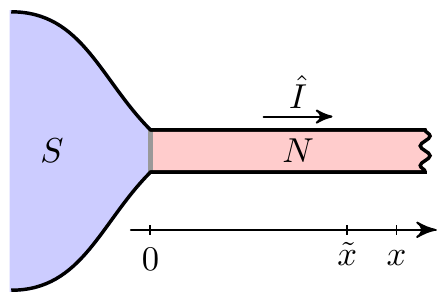}
\caption{A normal wire attached to a superconducting terminal.}
\label{sn-fig}
\end{figure}

Owing to the absence of quasiparticle states in both terminals $S_{1}$ and $S_{2}$ at subgap energies the matrix conductance of an SN interface acquires a particularly simple structure at such energies. Let $\check G_-$ be the bulk Green function for a superconducting terminal. Employing the condition $\hat G_-^R = \hat G_-^A$ applicable at subgap energies we observe from Eqs. \eqref{jeNazarov}, \eqref{gLNazarov} and \eqref{gMNazarov} that $g^L$-component of the matrix conductance vanishes identically, while its $g^{\mathcal{Y}}$-component is linked to the spectral current $j_{\varepsilon}$ by means of a simple formula
\begin{equation}
j_{\varepsilon} =  \dfrac{2}{\mathcal{A} \sigma} g^{\mathcal{Y}}.
\end{equation}
Making use of the above relations we conclude that that at subgap energies the matrix conductance for the SN interface can be parameterized by the two spectral conductances $g^T$ and $G_j$, i.e.
\begin{equation}
\hat G_{-+}
=
\begin{pmatrix}
-g^T & 0 \\
G_j & 0
\end{pmatrix}
, 
\quad
\hat G_{+-}
=
\begin{pmatrix}
g^T & G_j \\
0 & 0
\end{pmatrix}.
\label{hatGSNNazarov}
\end{equation}
Considering now a normal wire connected to the superconducting terminal via some barrier (Fig. \ref{sn-fig}) and employing the Kirchhoff rule for series resistances derived in Appendix A one can evaluate the effective subgap matrix conductance for a complex resistor consisting of both the interface barrier and the attached normal wire. After a simple calculation we obtain
\begin{equation}
\hat G_{0,x}
=
\begin{pmatrix}
-G & 0 \\
G_j & 0
\end{pmatrix}
, 
\quad
\hat G_{x,0}
=
\begin{pmatrix}
G & G_j \\
0 & 0
\end{pmatrix},
\label{hatGSN}
\end{equation}
where $G(\varepsilon)$ is a spectral parameter characterizing both the interface and the normal wire. Note that the above structure of the subgap matrix conductance is by no means accidental being consistent with the fundamental observation stating that the heat spectral current $\varepsilon I^L$ vanishes identically at subgap energies.

For fully transparent interfaces the boundary conditions \eqref{BcNaz} reduce to a simple continuity condition for the Green functions across the interface which reads $\check G_- = \check G_+$. In this case the functions $\mathcal{Y}$ and $D^L$ vanish at the SN interface and, hence, the matrix $\hat D$ becomes singular at this interface. Hence, special care should be taken while evaluating the subgap matrix conductance of the normal wire attached directly to the superconducting electrode. In Appendix \ref{ASN} we demonstrate that the matrix conductance structure in Eq. \eqref{hatGSN} remains preserved also in this particular case.

\section{Thermoelectric and Josephson effects}

Let us now employ the above circuit theory formalism in order to describe phase-coherent thermoelectric an Josephson effects in Andreev interferometers displayed in Fig. \ref{snns3-fig}. 

As we already indicated above, keeping the normal terminals $N_1$ and $N_2$ at different temperatures $T_1$ and $T_2$ one drives the quasiparticle distribution functions inside our structure out of equilibrium which in general makes the whole problem rather difficult to deal with. Some simplifications can be achieved provided we assume that the order parameter value $|\Delta |$ in both superconducting terminals strongly exceeds both temperatures $T_{1,2}$ and thermoelectric voltages $eV_{1,2}$ as well as the characteristic Thouless energy of our structure $E_{\rm Th}=D/L^2$ (where $L=L_p+L_{S_1}+L_{S_2}$) along with temperature $T$ of both superconducting terminals. Except pointed out otherwise, below we will adopt this assumption which allows to disregard the effect of overgap quasiparticles and focus our attention only on subgap electron transport.

Further drastic simplifications are achieved if we make use of Eq. \eqref{kirmat} which allows us to write down the relations between the matrix currents and the matrix distribution functions for each of the five wires in our structure. We obtain
\begin{gather}
\begin{pmatrix}
G_{S_1} & G_j \\
0 & 0
\end{pmatrix}
\hat H_{p_1} = -e \hat I_{S_1},
\label{IS1}
\\
\begin{pmatrix}
G_{S_2} & -G_j \\
0 & 0
\end{pmatrix}
\hat H_{p_2} = -e \hat I_{S_2},
\label{IS2}
\\
\hat G_{N_1}
(\hat H_{N_1} - \hat H_{p_1})
= e \hat I_{N_1},
\label{IN1}
\\
\hat G_{N_2}
(\hat H_{N_2} - \hat H_{p_2})
= e \hat I_{N_2},
\label{IN2}
\\
\hat G_{p_1 p_2} \hat H_{p_1} +  \hat G_{p_2 p_1} \hat H_{p_2} = -e \hat I_p,
\label{Ip1}
\end{gather}
where $\hat H_{p_1}$ and $\hat H_{p_2}$ are the matrix distribution functions at the nodes $p_1$ and $p_2$ and the spectral matrix 
conductances $\hat G_{N_{1,2}}$, $\hat G_{p_1 p_2}$ and $\hat G_{p_2 p_1}$ can be expressed in the form
\begin{gather}
\hat G_{N_{1,2}}
=
\begin{pmatrix}
G_{N_{1,2}}^T & G_{N_{1,2}}^{\mathcal{Y}} \\
- G_{N_{1,2}}^T & G_{N_{1,2}}^L
\end{pmatrix}
\\
\hat G_{p_1 p_2}
=
-
\begin{pmatrix}
G_p^T & G_p^{\mathcal{Y}} - G_j/2
\\
- G_p^{\mathcal{Y}} - G_j/2 & G_p^L
\end{pmatrix},
\\
\hat G_{p_2 p_1}
=
\begin{pmatrix}
G_p^T & G_p^{\mathcal{Y}} + G_j/2
\\
- G_p^{\mathcal{Y}} + G_j/2 & G_p^L
\end{pmatrix}.
\end{gather}

The distribution functions $\hat H_{N_{1,2}}$ inside the normal terminals read  
\begin{gather}
\hat H_{N_{1,2}} = 
\begin{pmatrix}
h^T_{N_{1,2}} \\ h^L_{N_{1,2}}
\end{pmatrix},
\\
\label{hT12}
h^T_{N_{1,2}}
=
\dfrac{1}{2}
\left[
\tanh\dfrac{\varepsilon + eV_{1,2} }{2 T_{1,2}} -
\tanh\dfrac{\varepsilon - eV_{1,2} }{2 T_{1,2}}
\right],
\\
\label{hL12}
h^L_{N_{1,2}}
=
\dfrac{1}{2}
\left[
\tanh\dfrac{\varepsilon + eV_{1,2} }{2 T_{1,2}} +
\tanh\dfrac{\varepsilon - eV_{1,2} }{2 T_{1,2}}
\right].
\end{gather}

In addition, it is necessary to employ the continuity conditions \eqref{curcons} for the matrix current at the crossing points $p_1$ and $p_2$, i.e.
\begin{gather}
\hat I_{S_1} + \hat I_{N_1} = \hat I_p,
\quad
\hat I_{S_2} + \hat I_{N_2} + \hat I_p = 0.
\label{Ip2}
\end{gather}

Making use of the above system of linear equations one can evaluate the matrix distribution functions $\hat H_{p_1}$ and $\hat H_{p_2}$ as well as the matrix currents depending on the distribution functions $h^{T,L}_{N_{1,2}}$ in the normal terminals. Then one can derive the general expressions for thermoelectric voltages $V_1$, $V_2$ along with the supercurrent $I_S$ as functions of the phase difference $\chi = \chi_1-\chi_2$ between the superconducting electrodes, temperatures $T_1$ and $T_2$ and other relevant parameters. 

\subsection{Symmetric structures}
Let us focus our attention on a special case of symmetric interferometers in which case the whole analysis becomes simpler due to the presence of extra symmetry conditions. Setting $L_{N_1}=L_{N_2}\equiv L_{N}$, $L_{S_1}=L_{S_2}\equiv L_{S}$, $\mathcal{A}_{N_1} = \mathcal{A}_{N_1}\equiv \mathcal{A}_N$, and $\mathcal{A}_{S_1} = \mathcal{A}_{S_1}\equiv \mathcal{A}_S$ we gain extra relations between spectral conductances for different wire segments: 
\begin{gather}
G_{S_1} = G_{S_2} = G_S,
\\
G_{N_{1,2}}^{T,L} = G_N^{T,L},
\\
G_{N_1}^{\mathcal{Y}} = - G_{N_2}^{\mathcal{Y}} = G_N^{\mathcal{Y}},
\quad
G_p^{\mathcal{Y}} = 0.
\end{gather}
With the aid of these relation it becomes possible to decouple the equations for the combinations of the distribution functions $\hat H_{p_1} \pm \hat \tau_3 \hat H_{p_2}$ and write
\begin{multline}
\begin{pmatrix}
G_S + G_N^T & G_N^{\mathcal{Y}} + G_j \\
- G_N^{\mathcal{Y}} - G_j & G_N^L + 2 G_p^L
\end{pmatrix}
(\hat H_{p_1} + \hat \tau_3 \hat H_{p_2})
\\=
\hat G_N (\hat H_{N_1}  + \hat \tau_3 \hat H_{N_2}),
\label{HN1sym2}
\end{multline}
and 
\begin{multline}
\begin{pmatrix}
G_S + G_N^T + 2 G_p^T & G_N^{\mathcal{Y}}  \\
- G_N^{\mathcal{Y}} & G_N^L
\end{pmatrix}
(\hat H_{p_1} - \hat \tau_3 \hat H_{p_2})
\\=
\hat G_N (\hat H_{N_1} - \hat \tau_3 \hat H_{N_2}).
\label{HN2sym2}
\end{multline}
These equations can easily be resolved providing the expressions for $\hat H_{p_1}$ and $\hat H_{p_2}$ in terms of the terminal distribution functions $\hat H_{N_1}$ and $\hat H_{N_2}$. 

The conditions \eqref{zcc} yield
\begin{gather}
\int
\dfrac{
(G_S + 2 G_p^T) [G_N^L G_N^T + (G_N^{\mathcal{Y}})^2]
}{G_N^L (G_N^T + G_S + 2 G_p^T) + (G_N^{\mathcal{Y}})^2
}
(h_{N_1}^T - h_{N_2}^T)
d \varepsilon =0,
\label{eqV1}
\end{gather}
and
\begin{widetext}
\begin{multline}
\int 
\dfrac{G_N^T (2 G_S G_p^L + G_j^2) + G_S [G_N^T G_N^L + (G_N^{\mathcal{Y}})^2]}{
(G_S + G_N^T) (2 G_p^L + G_N^L) + (G_j + G_N^{\mathcal{Y}})^2
}
(h_{N_1}^T + h_{N_2}^T)
d \varepsilon
\\+
\int 
\dfrac{ G_N^{\mathcal{Y}} (G_S 2 G_p^L + G_j^2) + G_j [G_N^T G_N^L + (G_N^{\mathcal{Y}})^2] 
}{
(G_S + G_N^T) (2 G_p^L + G_N^L) + (G_j + G_N^{\mathcal{Y}})^2
}
(h_{N_1}^L - h_{N_2}^L)
d \varepsilon
=0.
\label{eqV2}
\end{multline}
The general expression for the supercurrent $I_S$ can be derived from Eq. \eqref{IS1} together with matrix distribution function $\hat H_{p_1}$ evaluated from Eqs. \eqref{HN1sym2} and \eqref{HN2sym2}. We obtain
\begin{equation}
eI_S
=
-
\dfrac{1}{4}
\int
\left\{
\dfrac{G_S [ G_N^T G_N^L + (G_N^{\mathcal{Y}})^2] - G_j G_N^{\mathcal{Y}} (G_S + 2 G_p^T)
}{G_N^L (G_N^T + G_S + 2 G_p^T) + (G_N^{\mathcal{Y}})^2}
(h_{N_1}^T - h_{N_2}^T)
+
G_j (h_{N_1}^L + h_{N_2}^L)
\right\}
d \varepsilon.
\label{ISsym}
\end{equation}
\end{widetext}
Let us emphasize that Eqs. \eqref{eqV1}-\eqref{ISsym} involve \textit{no approximations} and represent a full solution of the problem of the subgap electron transport in symmetric Andreev interferometers displayed in Fig. \ref{snns3-fig}. Equations \eqref{eqV1}-\eqref{eqV2} allow to evaluate the thermoelectric voltages $V_{1,2}$ induced in the normal terminals, while Eq. \eqref{ISsym} determines the Josephson current flowing between the superconducting terminals $S_1$ and $S_2$ in the presence of a temperature gradient applied to normal ones $N_1$ and $N_2$.


Below we will specifically address the limits of sufficiently high and low temperatures and explicitly resolve  Eqs. \eqref{eqV1}-\eqref{ISsym}  in these two limits.

\subsection{High temperature limit}
We first consider the high temperature limit $T_{1,2} \gg E_{\mathrm{Th}}$. It is easy to observe that in this limit energies $|\varepsilon| \sim T_{1,2}$ provide the main contribution to the integral in  Eq. \eqref{eqV1}. In this case the spectral conductances $G_N^{\mathcal{Y}}$ are exponentially small and, hence, can be disregarded, the conductances $G_p^T$ and $G_N^{T}$ in Eq. \eqref{eqV1} can simply be replaced by their normal state values, respectively $G_p^n$ and $G_N^n$,
and, finally, the conductance $G_S$ can be taken in the form \cite{BWBSZ1999} 
\begin{equation}
G_S \simeq G_S^n\left(1+ \dfrac{\alpha}{L_S} \sqrt{\frac{D}{|\varepsilon |}}\right),\quad \alpha \simeq 0.29.
\label{GS}
\end{equation}
where $G_S^n$ is the corresponding normal state conductance. Performing all these manipulations in Eq. \eqref{eqV1}, combining it with Eqs. \eqref{hT12}, introducing the voltages $V=(V_1+V_2)/2$, $\delta V= V_1 - V_2$ and making use of the inequality $|\delta V| \ll |V|$ that holds in the high temperature limit considered here, one arrives
the following relation between $V$ and $\delta V$:
\begin{multline}
\frac{\delta V}{V}= -\dfrac{\alpha {\mathcal I} G_N^n G_S^n }{ (G_N^n + G_S^n + 2 G_p^n) (G_S^n + 2 G_p^n)}
\\\times
 \dfrac{L}{L_S} \sqrt{\dfrac{E_{\mathrm{Th}}}{2}}
\left(
\dfrac{1}{\sqrt{T_1}}
-
\dfrac{1}{\sqrt{T_2}}
\right),
\label{V1-V2}
\end{multline}
where 
\begin{equation}
{\mathcal I}=\int\limits_0^{\infty}\dfrac{ dx}{\sqrt{x}\cosh^2 x} =
\dfrac{2 (1 - 1/\sqrt{8})}{\sqrt{\pi}}\zeta(3/2)
\approx 1.9.
\end{equation}

In order to evaluate the thermoelectric voltage $V$ it is necessary to employ Eq. \eqref{eqV2} combined with Eqs. \eqref{hT12} and \eqref{hL12}. In the leading order in the parameter  $E_{\mathrm{Th}}/T_{1,2}\ll 1$ it suffices to replace the combination in front of $h_{N_1}^T + h_{N_2}^T$ in Eq. \eqref{eqV2} by its normal state value. Then after simple algebra we obtain
\begin{widetext}
\begin{gather}
V
=
\dfrac{1}{2}
\dfrac{G_N^n
}{
G_S^n (2 G_p^n + G_N^n)
}
[I_J(T_1,\chi) - I_J(T_2,\chi)]
+
\left(
\dfrac{1}{T_1} - \dfrac{1}{T_2}
\right)
\int [K_{\mathcal{Y}}(\varepsilon, \chi) + K_j(\varepsilon, \chi)]\varepsilon d \varepsilon,
\label{Vsym}
\\
K_{\mathcal{Y}}(\varepsilon, \chi) = -
\dfrac{G_S^n + G_N^n}{8eG_S^n G_N^n}
G_N^{\mathcal{Y}} 
\dfrac{ (2 G_S G_p^L + G_j^2)
}{
(G_S + G_N^T) (2 G_p^L + G_N^L) + (G_j + G_N^{\mathcal{Y}})^2
},
\label{KM}
\\
K_j(\varepsilon, \chi) = -
\dfrac{G_j }{8eG_S^n }
\Biggl[
\dfrac{ G_N^T G_N^L + (G_N^{\mathcal{Y}})^2 
}{
(G_S + G_N^T) (2 G_p^L + G_N^L) + (G_j + G_N^{\mathcal{Y}})^2
}
\dfrac{G_S^n + G_N^n}{G_N^n }
-
\dfrac{G_N^n
}{
(2 G_p^n + G_N^n)
}
\Biggr],
\label{Kj}
\end{gather}
\end{widetext}
where $I_J(T, \chi)$ is the equilibrium Josephson current in our structure at temperature $T$. Note that the first term in Eq. \eqref{Vsym}
proportional to the difference of the Josephson currents  $I_J(T_1,\chi) - I_J(T_2,\chi)$ agrees with that previously derived by Virtanen and  
Heikkil\"a \cite{VH} whereas the last term $\propto  1/T_1 - 1/T_2$ does not coincide with the corresponding extra contribution to $V$ found in that work. We observe that the kernel $K_{\mathcal{Y}}$ \eqref{KM} is fully determined by electron-hole asymmetry in the spectrum, and it vanishes identically provided this asymmetry is absent. In contrast, the kernel $K_j$ \eqref{Kj} has a mixed origin and remains non-zero even if electron-hole symmetry would be restored.

Now let us specify the expression for the Josephson current $I_S$ in the presence of a temperature gradient $T_1-T_2$. In the limit $T_{1,2} \gg E_{\mathrm{Th}}$ Eq. \eqref{ISsym} reduces to
\begin{widetext}
\begin{equation}
I_S
=
\dfrac{1}{2}
[I_J(T_1, \chi) + I_J(T_2, \chi)]
-
\dfrac{1}{4e}
\int
\dfrac{G_S [ G_N^T G_N^L + (G_N^{\mathcal{Y}})^2] - G_j G_N^{\mathcal{Y}} (G_S + 2 G_p^T)
}{G_N^L (G_N^T + G_S + 2 G_p^T) + (G_N^{\mathcal{Y}})^2}
(h_{N_1}^T - h_{N_2}^T)
d \varepsilon .
\label{IShT}
\end{equation}
\end{widetext}
The last term in  Eq. \eqref{IShT} can be significantly simplified making use of Eq. \eqref{eqV1}. With this in mind after some algebraic manipulations from Eq. \eqref{IShT} we obtain
\begin{equation}
I_S=\dfrac{1}{2}[I_J(T_1, \chi) + I_J(T_2, \chi)]+G_p^n \delta V.
\label{ISsym2}
\end{equation}

Equations \eqref{V1-V2}, \eqref{Vsym} and \eqref{ISsym2} provide the expressions both for thermoelectric voltages $V$ and $\delta V$ and for the Josephson current $I_S$ which are formally exact in the high temperature limit $T_{1,2} \gg E_{\mathrm{Th}}$. They also illustrate an intimate relation between thermoelectric and Josephson effects in the presence of a temperature gradient. 

In the above expressions one can identify the two types of terms, quasi-equilibrium and non-equilibrium ones. Quasi-equilibrium terms contain the combinations $I_J(T_1) \pm I_J(T_2)$ which decay exponentially at temperatures exceeding  $E_{\mathrm{Th}}$. Thus, at
sufficiently high values of $T_{1,2}$ these quasi-equilibrium terms can be safely neglected and the expressions for $V$, $\delta V$ and $I_S$ will be dominated by non-equilibrium ones which -- according to Eqs. \eqref{V1-V2}, \eqref{Vsym} and \eqref{ISsym2} -- yield slower (power law) temperature dependencies, i.e.
\begin{equation}
V \propto \dfrac{1}{T_1} - \dfrac{1}{T_2}
\end{equation}
and
\begin{equation}
\delta V \propto I_S \propto \left(\dfrac{1}{T_1} - \dfrac{1}{T_2}\right)\left(
\dfrac{1}{\sqrt{T_2}}
-
\dfrac{1}{\sqrt{T_1}}
\right).
\label{dVIS}
\end{equation}
We also note that in the limit $L_p \to 0$ both thermoelectric voltages $V$ and $\delta V$ vanish identically together with the non-equilibrium contribution to the supercurrent $I_S$. These observations fully agree with the results derived earlier for the case of symmetric $X$-junctions \cite{KDZ20}. 

In order to specify the quasi-equilibrium contributions to both thermoelectric voltages and the Josephson current it suffices to simply evaluate the equilibrium supercurrent $I_J$ at a given temperature $T\gg E_{\mathrm{Th}}$. 
This task can easily be accomplished (see Appendix \ref{Ffunctions}) with the result
\begin{multline}
I_J(T,\chi)
=\frac{128}{3+2\sqrt{2}}
\dfrac{E_\mathrm{Th}}{eL}
\dfrac{\mathcal{A}_S^2 \mathcal{A}_p \sigma}{
(\mathcal{A}_S + \mathcal{A}_N + \mathcal{A}_p)^2}
\sin \chi
\\\times
\left(\dfrac{2 \pi T}{E_\mathrm{Th}}\right)^{3/2}
e^{-\sqrt{2 \pi T/E_\mathrm{Th}} },
\label{IJ2}
\end{multline}
which reduces to the standard expression for the supercurrent in SNS junctions \cite{ZZh,GreKa} provided we set  
$\mathcal{A}_p=\mathcal{A}_S$ and $\mathcal{A}_N\to 0$. 

The remaining non-equilibrium terms originating from the last term in Eq. \eqref{Vsym} contain certain combinations of the spectral conductances \eqref{KM}, \eqref{Kj} integrated over energy. Hence, explicit evaluation of such terms requires solving the Usadel equations 
at all energy values. While at small and large energies (as compared to the Thouless energy) this problem can be handled analytically,
for energies  $ |\varepsilon| \sim E_{\mathrm{Th}}$ only a numerical solution is possible. In order to proceed we first evaluate the non-equilibrium terms in Eqs. \eqref{V1-V2}, \eqref{Vsym} and \eqref{ISsym2} approximately employing analytic results for the spectral conductances derived at higher energies $ |\varepsilon| >  E_{\mathrm{Th}}$ and then combine these approximate results with a numerically exact calculation.

Notice that in the interesting for us high energy limit $ |\varepsilon| \gg E_{\mathrm{Th}}$ the kernels  \eqref{KM} and \eqref{Kj} can be significantly simplified. In this limit we can neglect higher powers of $G_j$ and $G^{\mathcal{Y}}$ conductances and replace $G^{T,L}_{N,p}$ by its normal state values. Moreover, $G_S$ can be replaced by $G_S^n$ in $K_{\mathcal{Y}}$ kernel whereas in $K_j$ kernel it is necessary to keep correction term for $G_S$. As a result we obtain
\begin{gather}
K_{\mathcal{Y}} \simeq - 
\dfrac{G_N^{\mathcal{Y}}}{4e (2 G_p^n + G_N^n)}
\dfrac{G_p^n}{G_N^n},
\label{KMapprox}
\\
K_j \simeq -
\dfrac{G_N^nG_j 
(G_S^n - G_S)}{8eG_S^n (2 G_p^n + G_N^n) (G_S^n + G_N^n)}.
\label{Kjapprox}
\end{gather}
Here the expression for $G_S$ is defined in Eq. \eqref{GS}, whereas the conductances $G_j$ and $G_N^{\mathcal{Y}}$ can easily be recovered employing the solution of the Usadel equations at higher energies $ |\varepsilon| \gg E_{\mathrm{Th}}$ worked out in Appendix \ref{Ffunctions}, cf. Eqs. \eqref{Gjapprox} and \eqref{GMasym}. Substituting Eqs.  \eqref{GS}, \eqref{Gjapprox}, \eqref{GMasym} into Eqs. \eqref{KMapprox} and \eqref{Kjapprox}, combining the latter two equations with Eq. \eqref{Vsym} and formally extending the integral over $\varepsilon$ to all energies we arrive at the following result for the symmetric part of the induced thermoelectric voltage
\begin{multline}
V
=
\dfrac{1}{2}
\dfrac{G_N^n
}{
G_S^n (2 G_p^n + G_N^n)
}
[I_J(T_1,\chi) - I_J(T_2,\chi)]
\\+
\frac{32}{3+2\sqrt{2}}\dfrac{E^2_{\mathrm{Th}}\sin \chi}{e(2 G_p^n + G_N^n)}
\left(
\dfrac{1}{T_1} - \dfrac{1}{T_2}
\right)
\dfrac{ \sigma\mathcal{A}_S^2 \mathcal{A}_p}{
(\mathcal{A}_S + \mathcal{A}_N + \mathcal{A}_p)^2}
\\\times
\Biggl\{
\dfrac{4\mathcal{A}_p}{
(\mathcal{A}_S + \mathcal{A}_N + \mathcal{A}_p)}
\dfrac{L^4(3 L^2 - L_p^2)}{L_N (L^2 + L_p^2)^3 } 
-
 \dfrac{3 \alpha G_N^n}{L_S (G_S^n + G_N^n)}
\Biggr\},
\label{Vsymapprox2}
\end{multline}
where $I_J(T,\chi)$ is defined in Eq. \eqref{IJ}. The asymmetric part of the thermoelectric voltage $\delta V$ and the supercurrent $I_S$ are then defined respectively by Eqs. \eqref{V1-V2} and \eqref{ISsym2} combined with Eq. \eqref{Vsymapprox2}. 

From the above results we observe that at high temperatures $T_{1,2} \gg E_{\mathrm{Th}}$ symmetric Andreev interferometers exhibit purely sinusoidal dependence of both thermoelectric voltages $V$ and $\delta V$ as well as the supercurrent $I_S$ on the Josephson phase $\chi$, i.e.  $V\propto \delta V \propto I_S \propto \sin \chi$. Provided the non-equilibrium contribution $G_p^n \delta V$ in Eq. \eqref{ISsym2} exceeds the quasi-equilibrium one, the sign of the supercurrent $I_S$ is negative for any positive value of the combination in the curly brackets in 
Eq. \eqref{Vsymapprox2}. Hence, in this case the system is driven into a $\pi$-junction state, cf. also Eq. \eqref{dVIS}.

\begin{figure}
\begin{center}
\includegraphics[width=80mm]{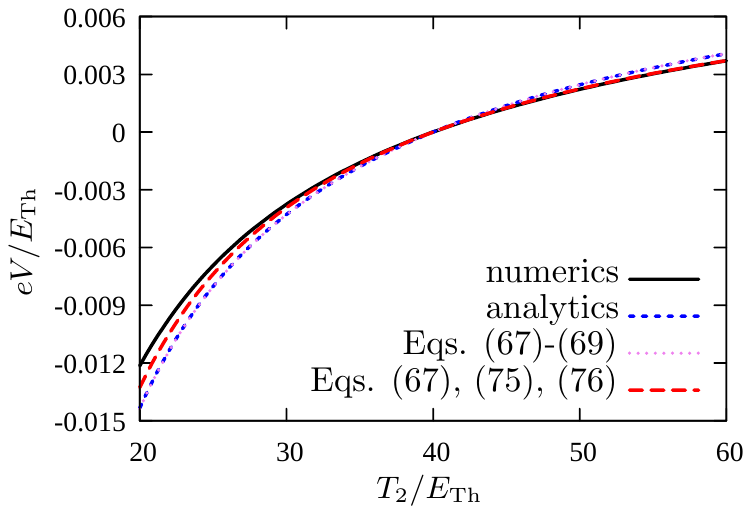}
\end{center}
\caption{Symmetric part of the thermoelectric voltage $V$ as a function of temperature $T_2$.  Solid line corresponds to a numerically exact solution of the Usadel equation, short dashed line indicates our analytic result in Eq. \eqref{Vsymapprox2}, dotted and long dashed lines represent Eq. \eqref{Vsym} combined
 with exact and approximate expressions for $K_{\mathcal{Y}}$ and $K_{j}$ defined respectively in Eqs. \eqref{KM}, \eqref{Kj} and in Eqs. \eqref{KMapprox}, \eqref{Kjapprox}.  The chosen parameter values $T_1=40 E_{\mathrm{Th}}$, $\chi=\pi/2$, $L_N=L_S=L_p=L/3$  and $\mathcal{A}_S = \mathcal{A}_N = \mathcal{A}_p$ remain the same for all curves.}
\label{V-T-033-T1-40-3-fig}
\end{figure}

\begin{figure}
\begin{center}
\includegraphics[width=80mm]{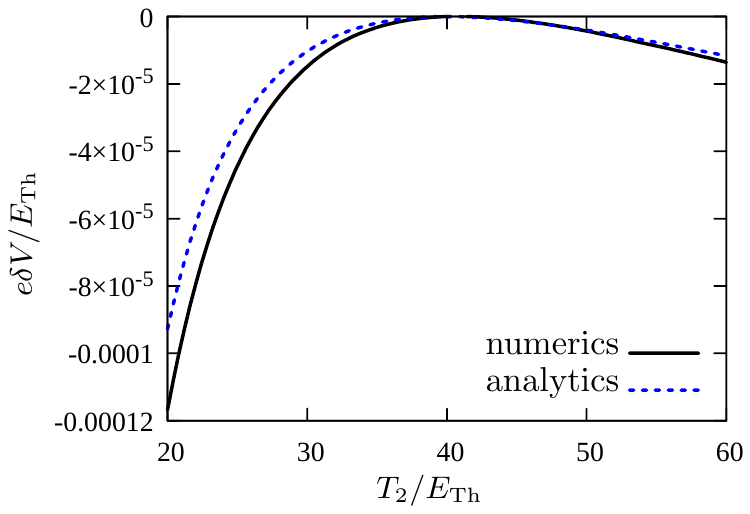}
\end{center}
\caption{Asymmetric part of the thermoelectric voltage $\delta V$ as a function of $T_2$.  Solid line corresponds to a numerically exact solution of the Usadel equation, dashed line indicates our analytic result defined in Eq. \eqref{V1-V2} combined with Eq. \eqref{Vsymapprox2}.  The parameter values are the same as in Fig. \ref{V-T-033-T1-40-3-fig}.}
\label{dV-T-033-T1-40-fig}
\end{figure}

In order to verify the accuracy of the employed simple approximations we also evaluated the thermoelectric voltages $V$ and $\delta V$ numerically by directly solving the Usadel equations. The corresponding results are displayed in Figs. \ref{V-T-033-T1-40-3-fig} and \ref{dV-T-033-T1-40-fig} as functions of temperature for one of the two normal terminals together with our analytic estimates for  $V$ and $\delta V$. For both these quantities we observe a remarkably good agreement between numerically exact results and those defined by our Eqs. \eqref{Vsymapprox2} and \eqref{V1-V2}.

Note that the first and the second terms in curly brackets in Eq. \eqref{Vsymapprox2} originate respectively from $K_{\mathcal{Y}}$- and $K_j$-terms. We verified that our simple approximation \eqref{KMapprox} for $K_{\mathcal{Y}}$ remains sufficiently accurate in a wide range of parameters of our structure. At the same time, the approximation \eqref{Kjapprox} for $K_j$ may sometimes become less accurate since the high energy expansion in Eq. \eqref{GS} is not supposed to work well at lower energies $|\varepsilon| \lesssim E_{\mathrm{Th}}$. 

\subsection{Lower temperatures}
Let us now briefly address the opposite low temperature limit $T_{1,2} \ll E_{\mathrm{Th}}$. In this limit the integrals in Eqs. \eqref{eqV1}-\eqref{eqV2} are restricted to energy intervals where all spectral conductances behave as smooth functions of energy and with a good accuracy can be expanded in Taylor series near $\varepsilon =0$. Below we will also make use of the fact that diagonal elements of the matrix conductances are even functions of $\varepsilon$, whereas their off-diagonal elements are odd functions of energy.

At temperatures well below the Thouless energy $E_{\mathrm{Th}} $ the combination of spectral conductances in front of the term $(h_{N_1}^T - h_{N_2}^T)$ under the energy integral in Eq. \eqref{eqV1} can be simplified and replaced by
\begin{equation}
\dfrac{
(G_S^n + 2 G_p^n) G_N^n 
}{ (G_N^n + G_S^n + 2 G_p^n) 
}
[1 + c (\varepsilon /E_{\mathrm{Th}})^2 ], \quad c \sim 1.
\end{equation}
Then Eq. \eqref{eqV1} yields
\begin{equation}
\frac{\delta V}{V} = - \dfrac{\pi^2c}{3} \dfrac{T_1^2 - T_2^2}{E^2_{\mathrm{Th}}} ,
\label{dVtheorshort}
\end{equation}
i.e. for symmetric interferometers the inequality  $|\delta V| \ll |V|$ holds also in the low temperature limit $T_{1,2} \ll E_{\mathrm{Th}}$. 

As before, the symmetric part of the thermoelectric voltage $V$ can be derived from Eq. \eqref{eqV2}. Making use of the fact that 
the spectral conductances $G_S$ and $G^T_{N,p}$ evaluated in the zero energy limit exactly coincide with their normal state values due to the so-called reentrance effect \cite{BWBSZ1999,Nazarov96,golubov1997coherent} we obtain
\begin{multline}
V
=
\dfrac{\pi^2}{12}
\dfrac{ (G_N^{\mathcal{Y}})' G_S^n 2 G_p^L(0) + G_j' G_N^n G_N^L(0)
}{
eG_S^n G_N^n [2 G_p^L(0) + G_N^L(0)]
}
(T_1^2 - T_2^2).
\label{Vtheorshort}
\end{multline}
Here and below $G^L_{p,N}(0)$ define the spectral conductances in the zero energy limit while $(G_N^{\mathcal{Y}})'$ and $G_j'$ represent the derivatives of the corresponding spectral conductances with respect to energy  at $\varepsilon=0$. For the supercurrent $I_S$ we get
\begin{equation}
I_S
=
\dfrac{1}{2}
[I_J(T_1, \chi) + I_J(T_2, \chi)]
+
\dfrac{\pi^2}{6}V
\dfrac{T_1^2 - T_2^2}{E_{\mathrm{Th}}^2}
\tilde G,
\label{ISlT}
\end{equation}
where $V$ and $\tilde G$ are defined respectively in Eqs. \eqref{Vtheorshort} and \eqref{ISshort}. Having in mind that at low $T$ the temperature dependence of the equilibrium supercurrent is determined by the expression
\begin{equation}
I_J(T, \chi)  \simeq I_J(0, \chi) +\dfrac{\pi^2}{6} G_j' T^2, \quad G_j' < 0,
\label{JT}
\end{equation}
we conclude that in this limit the leading $T$-dependent correction to the supercurrent behaves as $\sim -(T_1^2+T_2^2)/E_{\mathrm{Th}}^2$ and originates from the quasi-equilibrium contribution to $I_S$. At the same time, the non-equilibrium terms produce only a subleading correction  $\sim (T_1^2-T_2^2)^2/E_{\mathrm{Th}}^4$ which can be safely neglected in the low temperature limit. 

Qualitatively the same conclusion holds also in the limit $T_1 \ll E_{\mathrm{Th}} \ll T_2$ in which case $I_S$ is again dominated by
the quasi-equilibrium contribution $I_S \simeq I_J(T_1, \chi) /2$.

Let us also remind the reader that everywhere in the above analysis we merely addressed the most relevant limit of not too short metallic wires with $E_{\mathrm{Th}} \ll |\Delta|$. It is easy to demonstrate that the latter condition can actually be relaxed while Eqs. \eqref{dVtheorshort}, \eqref{Vtheorshort} and \eqref{ISlT} will remain applicable also at $E_{\mathrm{Th}} \gtrsim |\Delta|$ (though still for $T_{1,2},T \ll |\Delta|$). 
In this case one should be somewhat more cautious since the contribution of quasiparticles with overgap energies should also be taken into account. Fortunately, however, at $|\varepsilon | \gg T_{1,2},T$ the spectral current $I^T$ depends neither on temperature nor on thermoelectric  voltages $V_{1,2}$, i.e. $eI^T = - G_j \sgn \varepsilon$. Combining this expression for $I^T$ with Eq. \eqref{ISsym} we again arrive at Eq. \eqref{ISshort}.

\begin{figure}
\begin{center}
\includegraphics[width=80mm]{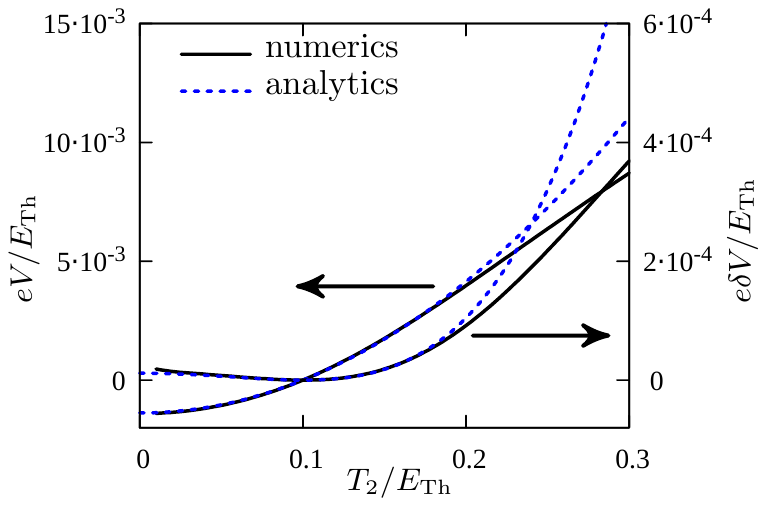}
\end{center}
\caption{Thermoelectric voltages $V$ and $\delta V$ as functions of temperature $T_2$ for $T_1 = T =0.1 E_{\mathrm{Th}}$, $\chi=\pi/2$, $|\Delta| = E_{\mathrm{Th}}$,  $L_N=L_S=L_p=L/3$,  and $\mathcal{A}_S = \mathcal{A}_N = \mathcal{A}_p$. Solid lines correspond to numerically exact solutions of the Usadel equation while dashed lines display analytic results for $V$ \eqref{Vtheorshort} and $\delta V$  \eqref{dVtheorshort}, where the prefactors are also evaluated numerically. }
\label{V-T-033-short-fig}
\end{figure}

In Fig. \ref{V-T-033-short-fig} we compare Eqs. \eqref{Vtheorshort} and \eqref{dVtheorshort} for the thermoelectric voltages $V$ and $\delta V$ as functions of $T_2$ with numerically exact results for these quantities obtained directly from the Usadel equation. We observe that at low temperatures our analytic results indeed remain very accurate and start deviating from numerical curves at $T_2 \gtrsim 0.2 E_{\mathrm{Th}}$.

\subsection{Asymmetric structures}

Our general approach can also be applied in order to describe both thermoelectric and non-equilibrium Josephson effects in asymmetric Andreev interferometers presented in Fig. \ref{snns3-fig}. The resulting expressions for $V_1$, $V_2$ and $I_S$ -- which provide a complete solution of our problem in this general case -- turn out to be rather lengthy and cumbersome. For that reason both the final results and their derivation are relegated to Appendix \ref{Aasym}. For symmetric structures these general results reduce to the corresponding expressions found above.

The behavior of asymmetric Andreev interferometers is in many respects qualitatively similar to that of asymmetric $X$-junctions  \cite{KDZ20,KZ20}. At high temperatures $T_{1,2} \gg E_{\mathrm{Th}}$ coherent thermoelectric voltages $V_{1}$ and $V_{2}$ generated at two normal metallic electrodes $N_1$ and $N_2$ in the presence of a temperature gradient are defined in Eqs. \eqref{V1-3} and \eqref{V2-3}. Similarly to the symmetric case, for asymmetric structures both thermoelectric voltages $V_{1,2}$ depend periodically on the phase difference $\chi$ and can be represented as a sum of a quasi-equilibrium contribution $\propto I_J(\chi,T_1) - I_J(\chi,T_2)$ and a non-equilibrium one being proportional to $1/T_1 - 1/T_2$ and dominating the result at higher temperatures. In contrast to symmetric structures, here the  voltage difference $\delta V= V_1 - V_2$ is not anymore small and can be of the same order as the voltage $V=(V_1+V_2)/2$ itself. 

The Josephson current $I_S$ in the limit $T_{1,2} \gg E_{\mathrm{Th}}$ is given by Eq. \eqref{ISasym} which also consists of quasi-equilibrium terms (defined as a sum of $I_J(\chi,T_1)$ and $I_J(\chi,T_2)$ "weighted" by certain combinations of normal state wire conductances) and non-equilibrium ones decaying as $\propto 1/T_1 - 1/T_2$. At high enough temperatures this non-equilibrium contribution exceeds the quasi-equilibrium one, thereby signaling about the possibility of a $\pi$-junction state. For strongly asymmetric structures and high temperatures the absolute value of $I_S$ can be rather large and may even reach the same order of magnitude as the equilibrium Josephson current at $T \to 0$, cf. \cite{KZ20}.

At low temperatures $T_{1,2} \ll E_{\mathrm{Th}} $ both quasi-equilibrium and non-equilibrium contributions to $V_1$ and $V_2$ are defined in Eqs. \eqref{V1theorshort}, \eqref{V2theorshort} and have the same temperature dependence $\propto T_1^2 - T_2^2$. The Josephson current 
$I_S$ is expressed by Eq. \eqref{Jas}.  Combining this result with Eq. \eqref{JT} we obtain $I_S\simeq I_J(0,\chi)$ minus small terms proportional to $T_1^2$ and $T_2^2$ associated with the temperature correction to the Josephson current \eqref{JT}
plus an extra term $\propto T_1^2 - T_2^2$ of a non-equilibrium origin. The latter term turns out to be parametrically larger than that for symmetric structures.

\section{Conclusions}

We worked out a detailed theory describing a non-trivial interplay between proximity-induced quantum coherence and non-equilibrium effects in Andreev interferometers exposed to an arbitrary temperature gradient.  

We elaborated a circuit theory applying it to the analysis of quantum coherent effects in the network of interconnected diffusive quasi-one-dimensional normal wires attached to external normal and superconducting terminals. We formulated transparent rules of the circuit theory 
resembling the standard Kirchhoff rules well known in the circuit electrodynamics. Our theory allows to explicitly derive the solution for the kinetic Usadel equation in terms of the spectral conductances. One of the key advantages of our approach is that it enables one to unambiguously identify different contributions responsible for a variety of physical phenomena involved in our problem. 

We demonstrated that the thermoelectric voltage response $V$ to an externally imposed temperature gradient $T_1-T_2$ depends periodically on the superconducting phase difference $\chi$ and is determined by the two groups of terms originating from different physical mechanisms. One of them is the quasi-equilibrium contribution proportional to the difference of the equilibrium Josephson currents $I_J(\chi,T_1) - I_J(\chi,T_2)$. This contribution exactly coincides with that identified previously \cite{VH}. It plays a significant role as long as at least one of the two temperatures -- $T_1$ or $T_2$ -- remains below an effective Thouless energy $E_{\mathrm{Th}}$ of our device. At the same time, this quasi-equilibrium contribution decays exponentially with increasing temperature and eventually becomes vanishingly small in the limit $T_{1,2} \gg E_{\mathrm{Th}}$. 

The thermoelectric response, however, does not vanish in this limit because of another contribution of essentially non-equilibrium origin. This non-equilibrium contribution involves different terms both related and unrelated to particle-hole asymmetry generated by the temperature gradient. As a result, with increasing temperature the thermoelectric voltage signal decays only as a power law $eV  \sim  E^2_{\mathrm{Th}}|1/T_1 - 1/T_2|$ and dominates the system behavior at high enough temperatures. We also note the thermoelectric response varies for two metallic terminals $N_1$ and $N_2$. At high temperatures the corresponding voltage difference $\delta V=V_1-V_2$ 
is parametrically smaller and decays faster than $V$ in the case of symmetric structures (cf. Eq. \eqref{dVIS}), whereas in asymmetric interferometers $\delta V$ gets larger and may reach the same order of magnitude as $V$.

We also investigated the Josephson current $I_S$ flowing across our hybrid structure between two superconducting terminals. 
Unlike in the equilibrium case, here both the magnitude of this current and its sign can be controlled not only by the phase difference $\chi$ and temperature, but also by the temperature gradient  $T_1-T_2$ applied to normal terminals of our device. Similarly to the thermoelectric signal, the Josephson current is determined by a sum of two different contributions -- quasi-equilibrium and non-equilibrium ones. The first contribution
takes a simple form $aI_J(\chi,T_1) + bI_J(\chi,T_2)$, where $a+b=1$ and, in particular, $a=b=1/2$ for symmetric interferometers.
The second -- non-equilibrium -- contribution gains importance at higher temperatures and dominates the supercurrent $I_S$ in the limit
$T_{1,2} \gg E_{\mathrm{Th}}$ where it decays as a power law with increasing $T_{1,2}$ in contrast to exponentially decaying equilibrium Josephson current. It is remarkable that in the presence of a temperature gradient $I_S$ can strongly exceed $I_J$ evaluated at any
of the two temperatures $T_1$ or $T_2$.

This supercurrent stimulation effect is directly related to the presence of non-equilibrium low-energy quasiparticles inside our device which suffer little dephasing while propagating between superconducting terminals.  Note that a somewhat similar situation is encountered for
the Aharonov-Bohm effect in superconducting-normal metallic heterostructures \cite{golubov1997coherent,courtois1996long} and for
the effect of supercurrent stimulation in long SNS junctions exposed to an external rf signal \cite{Aslamazov82,Zaikin83}. In both these cases long-range phase coherence is also maintained due to non-equilibrium quasiparticles with
energies below $E_{\mathrm{Th}}$ propagating inside the structure almost with no phase relaxation.

Yet another interesting phenomenon is the possibility of switching our device to the $\pi$-junction state by creating non-equilibrium
quasiparticles with the aid of a temperature gradient. Combining this effect with supercurrent stimulation one can realize a comparably large (cf., e.g., Eq. \eqref{1}) $\pi$-shifted Josephson current well observable in the limit $T_{1,2} \gg E_{\mathrm{Th}}$. Previously a similar Josephson current inversion effect was discussed in the context of driving the electron distribution function out of equilibrium by applying an external voltage bias $V_x$ to normal terminals \cite{WSZ,volkov1995,Yip,Teun}. Unlike here, however,
in that case the magnitude of the supercurrent remains exponentially small for $eV_x,T  \gg  E_{\mathrm{Th}}$ \cite{WSZ}. 
Supercurrent stimulation with an external voltage bias can nevertheless become possible in more complicated configurations where
non-equilibrium quasiparticles are supplied, e.g., by an extra normal terminal \cite{dolgirev2018interplay}.

Actually one can also combine both these physical situations by simultaneously exposing normal terminals to temperature gradient and voltage bias. In this case thermoelectric voltages $V_{1,2}$ should obey an obvious constraint $V_1 - V_2 = V_x$ and the current conservation condition $I_{N_1} + I_{N_2} = 0 $ should hold. Elaborating the circuit theory analysis developed here and observing these additional constraints it is straightforward to find both induced thermoelectric voltages $V_1$, $V_2$ and the Josephson current $I_S$ as functions of the phase difference $\chi$, temperatures $T_{1,2}$, bias voltage $V_x$ and other parameters. This task, however, goes beyond the frames of the present work.

The results presented here demonstrate that low temperature transport properties of superconducting hybrid nanostructures, such as Andreev interferometers, can be controlled and manipulated by means of a temperature gradient. This observation may create a variety of opportunities for novel applications of such structures in such fields as superconductivity-based quantum nanoelectronics, phase-coherent caloritronics and metrology.

\appendix

\section{Kirchhoff rules for matrix conductances}
\label{Aparallel}
\begin{figure}
\begin{center}
\includegraphics[width=80mm]{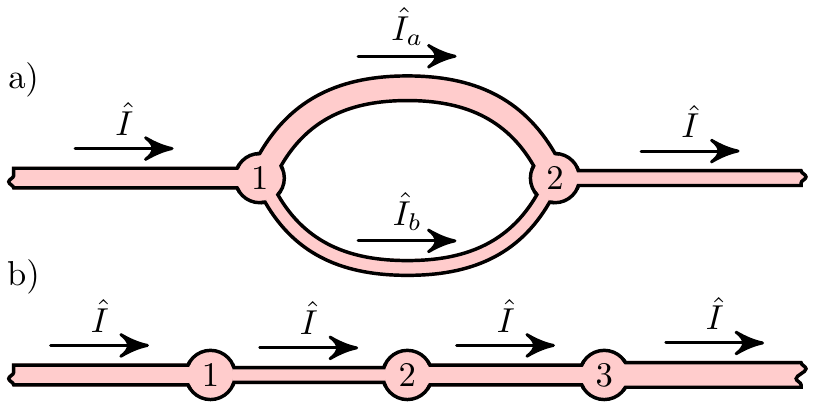}
\end{center}
\caption{Resistors connected in parallel (a) and in series (b).}
\label{parallel-series-fig}
\end{figure}
Let us establish simple Kirchhoff rules for matrix conductances of metallic wires connected either in parallel or in series.
Consider first a circuit displayed in Fig. \ref{parallel-series-fig}a. In this case the total matrix current $\hat I$ equals to a sum of the matrix currents in each of the two branches connected in parallel, i.e.  $\hat I = \hat I_a + \hat I_b$. Employing Eq. \eqref{kirmat} in both ``a'' and ``b'' branches
\begin{gather}
\hat G_{a,12} \hat H(1)
+
\hat G_{a,21} \hat H(2) = -e \hat I_a,
\\
\hat G_{b,12} \hat H(1)
+
\hat G_{b,21} \hat H(2) = -e \hat I_b,
\end{gather}
we immediately obtain
\begin{equation}
\hat G_{12} \hat H(1)
+
\hat G_{21} \hat H(2) = -e \hat I,
\end{equation}
where total conductances $\hat G_{12}$ and $\hat G_{21}$ read
\begin{gather}
\hat G_{12} = \hat G_{a,12} + \hat G_{b,12},
\quad
\hat G_{21} = \hat G_{a,21} + \hat G_{b,21}.
\end{gather}
We observe that these relations for matrix conductances are identical to those for equivalent conductance of two normal resistors connected in parallel.

In the case of the wires connected in series (see Fig. \ref{parallel-series-fig}b) Eq. \eqref{kirmat} can be written separately for each wire segment. We have
\begin{gather}
\hat G_{12} \hat H(1)
+
\hat G_{21} \hat H(2) = -e \hat I,
\\
\hat G_{23} \hat H(2)
+
\hat G_{32} \hat H(3) = -e \hat I,
\end{gather}
Combining these two equations, we obtain
\begin{equation}
\hat G_{13} \hat H(1)
+
\hat G_{31} \hat H(3) = -e \hat I,
\end{equation}
where the matrix conductances $\hat G_{13}$ and $\hat G_{31}$ read
\begin{gather}
\hat G_{13}^{-1} = \hat G_{12}^{-1} - \hat G_{12}^{-1} \hat G_{21} \hat G_{23}^{-1},
\label{series31}
\\
\hat G_{31}^{-1} = \hat G_{32}^{-1} - \hat G_{32}^{-1} \hat G_{23} \hat G_{21}^{-1}.
\label{series13}
\end{gather}
In the absence of superconductivity the matrix conductances are symmetric ($\hat G_{ij} = \hat G_{ji}$) being proportional to the unity matrix, cf.
Eq. \eqref{Rnorm}. In this limit Eqs. \eqref{series31}-\eqref{series13} simply reflect the standard rules for equivalent conductance of two normal resistors connected in series.

\section{Y-$\Delta$ transformation}
\label{AYD}
It is well known that an Y-shaped circuit composed of normal resistors can be transformed to a $\Delta$-shaped circuit of such resistors and vice versa.
\begin{figure}
\begin{center}
\includegraphics[width=80mm]{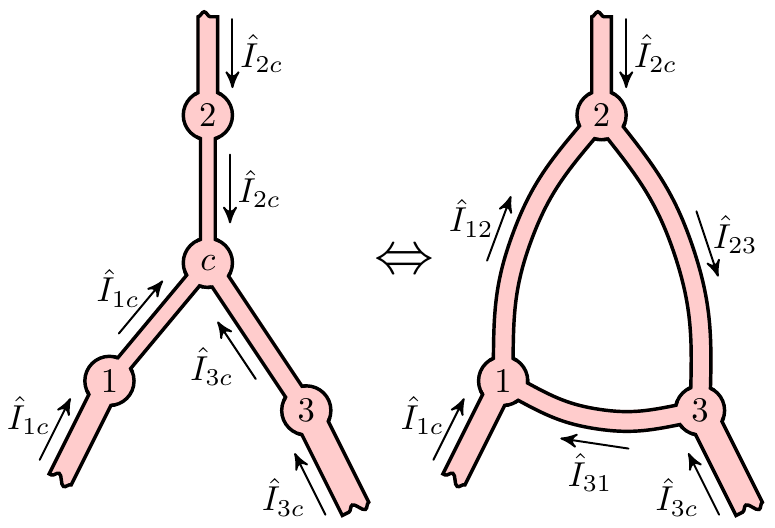}
\end{center}
\caption{Y-shaped and $\Delta$-shaped circuits.}
\label{delta-Y-fig}
\end{figure}
This Y-$\Delta$ transformation can be directly generalized to the case of matrix resistances considered here. Employing Eq. \eqref{kirmat} for the Y-shaped circuit displayed in the left part of Fig. \ref{delta-Y-fig}, we have
\begin{gather}
\hat G_{1c} \hat H_1 + \hat G_{c1} \hat H_c = - e \hat I_{1c},
\label{I1cY}
\\
\hat G_{2c} \hat H_2 + \hat G_{c2} \hat H_c = - e \hat I_{2c},
\label{I2cY}
\\
\hat G_{3c} \hat H_3 + \hat G_{c3} \hat H_c = - e \hat I_{3c}.
\label{I3cY}
\end{gather}
Making use of the matrix current conservation condition, we obtain an extra equation for the matrix currents $\hat I_{ic}$
\begin{equation}
\hat I_{1c} + \hat I_{2c} + \hat I_{3c} =0.
\end{equation}

In the case of the $\Delta$-shaped circuit displayed in the right part of Fig. \ref{delta-Y-fig} the relations between the matrix currents and the matrix distribution functions take the form 
\begin{gather}
\hat G_{12} \hat H_1 + \hat G_{21} \hat H_2 = - e \hat I_{12},
\label{I12D}
\\
\hat G_{23} \hat H_2 + \hat G_{32} \hat H_3 = - e \hat I_{23},
\label{I23D}
\\
\hat G_{31} \hat H_3 + \hat G_{13} \hat H_1 = - e \hat I_{31},
\label{I31D}
\end{gather}
whereas the matrix currents $\hat I_{ic}$ read 
\begin{gather}
\hat I_{1c} = \hat I_{12} - \hat I_{31},
\label{I1cD}
\\
\hat I_{2c} = \hat I_{23} - \hat I_{12},
\label{I2cD}
\\
\hat I_{3c} = \hat I_{31} - \hat I_{23}.
\label{I3cD}
\end{gather}
It is easy to verify that the relations between the matrix currents $\hat I_{ic}$ and the matrix distribution functions $\hat H_i$ for the $\Delta$-shaped circuit governed by Eqs. \eqref{I12D}-\eqref{I3cD} are equivalent to Eqs. \eqref{I1cY}-\eqref{I3cY} for the Y-shaped circuit provided we set
\begin{gather}
\hat G_{21} = -
\hat G_{c1} (\hat G_{c1} + \hat G_{c2} + \hat G_{c3})^{-1}\hat G_{2c},
\\
\hat G_{31} =
\hat G_{c1} (\hat G_{c1} + \hat G_{c2} + \hat G_{c3})^{-1}\hat G_{3c},
\\
\hat G_{12} = 
\hat G_{c2} (\hat G_{c1} + \hat G_{c2} + \hat G_{c3})^{-1}\hat G_{1c},
\\
\hat G_{32} = -
\hat G_{c2} (\hat G_{c1} + \hat G_{c2} + \hat G_{c3})^{-1}\hat G_{3c},
\\
\hat G_{13} =-
\hat G_{c3} (\hat G_{c1} + \hat G_{c2} + \hat G_{c3})^{-1}\hat G_{1c},
\\
\hat G_{23} =
\hat G_{c3} (\hat G_{c1} + \hat G_{c2} + \hat G_{c3})^{-1}\hat G_{2c},
\end{gather}

Note that the inverse transformation from the $\Delta$-shaped circuit to the Y-shaped one is not always possible. Under the condition
\begin{equation}
\hat G_{31}^{-1} \hat G_{21} \hat G_{23}^{-1} \hat G_{13} \hat G_{12}^{-1} \hat G_{32} = -1,
\end{equation}
this inverse transformation can be formulated. In this case it reads
\begin{gather}
\hat G_{1c} = \hat G_{12} - \hat G_{13} + \hat G_{21} \hat G_{23}^{-1} \hat G_{13},
\\
\hat G_{2c} = \hat G_{23} - \hat G_{21} + \hat G_{32} \hat G_{31}^{-1} \hat G_{21},
\\
\hat G_{3c} = \hat G_{31} - \hat G_{32} + \hat G_{13} \hat G_{12}^{-1} \hat G_{32},
\\
\hat G_{c1} = \hat G_{21} - \hat G_{31} + \hat G_{31} \hat G_{32}^{-1} \hat G_{12},
\\
\hat G_{c2} = \hat G_{32} - \hat G_{12} + \hat G_{12} \hat G_{13}^{-1} \hat G_{23},
\\
\hat G_{c3} = \hat G_{13} - \hat G_{23} + \hat G_{23} \hat G_{21}^{-1} \hat G_{31}.
\end{gather}

\section{Matrix conductance at subgap energies}
\label{ASN}
At subgap energies $|\varepsilon| < |\Delta|$ the kinetic equation \eqref{U} for the evolution operator $\hat U$ becomes singular in the vicinity of the interface between a superconducting terminal and a normal wire. This is because at such energies one has $\hat G^R= \hat G^A$ and, hence, the functions $D^L$  and $\mathcal{Y}$ vanish at the SN-interface together with the first derivative of $D^L$. 

In order to tackle this problem let us take a closer look at the solution of Eq. \eqref{U} in the vicinity of this interface. Expanding the kinetic coefficients $D^{L,T}$ and $\mathcal{Y}$ to the lowest nonvanishing order in the distance to SN-interface $x$ and making use of the identity $\hat G^R(0)= \hat G^A(0)$, we get
\begin{equation}
\hat D \approx
\begin{pmatrix}
d_T & j_{\varepsilon} x \\
- j_{\varepsilon} x & d_L x^2
\end{pmatrix},
\label{hatDSN}
\end{equation}
where $d_T = D^T(0)$, $d_L = {D^L}''(0)/2$ and $j_{\varepsilon} = \mathcal{Y}'(0)$. With this in mind the equation for the evolution operator can be solved explicitly, and we obtain
\begin{equation}
\hat U (x, \tilde x)
=
\dfrac{
\begin{pmatrix}
d_T d_L + j_{\varepsilon}^2 x / \tilde x & - d_L j_{\varepsilon} (x - \tilde x) \\
d_T j_{\varepsilon} (1/x - 1/\tilde x) & d_T d_L + j_{\varepsilon}^2 \tilde x / x
\end{pmatrix}
}{ d_T d_L + j_{\varepsilon}^2 }.
\end{equation}
The matrix conductance then reads
\begin{equation}
\hat G_{x, \tilde x} =
\dfrac{\sigma \mathcal{A}}{x - \tilde x}
\begin{pmatrix}
d_T & j_{\varepsilon} x \\
- j_{\varepsilon} \tilde x & d_L x \tilde x
\end{pmatrix}
\end{equation}
and, hence, we have
\begin{equation}
\hat G_{0, x} =
\dfrac{\sigma \mathcal{A}}{x}
\begin{pmatrix}
-d_T & 0 \\
j_{\varepsilon} x & 0
\end{pmatrix},
\quad
\hat G_{x, 0} =
\dfrac{\sigma \mathcal{A}}{x}
\begin{pmatrix}
d_T & j_{\varepsilon} x \\
0 & 0
\end{pmatrix}.
\end{equation}
Combining the above results with the relation for the two matrix conductances connected in series one can easily observe that the matrix conductance of the whole wire has the same structure as that in Eq. \eqref{hatGSN}.

\section{Equilibrium Josephson current and spectral conductances}
\label{Ffunctions}
In equilibrium the supercurrent $I_J$ flowing across our structure can be derived from the equation
\begin{equation}
I_J(T,\chi)= - \dfrac{\sigma \mathcal{A}}{2e}\int j_{\varepsilon} \tanh \dfrac{\varepsilon}{2T}
d \varepsilon,
\label{IJ}
\end{equation}
where $j_{\varepsilon}$ is the spectral current defined in Eq. \eqref{jE} and the cross-section $\mathcal{A}$ equals to either $\mathcal{A}_p$ or $\mathcal{A}_S$ depending on the wire in which $j_{\varepsilon}$ is evaluated. Accordingly, the task at hand is to evaluate
anomalous Green functions $F^R$ and $\tilde F^R$ inside all normal wires interconnecting superconducting and normal terminals. 

In general this task can be accomplished only numerically since quasiclassical Usadel equations constitute a complicated system of coupled nonlinear differential equations. However, in the high energy limit Usadel equations can be simplified and resolved analytically for an arbitrary network of quasi-one-dimensional normal wires connected to each other and to normal or superconducting terminals. 
\begin{figure}
\includegraphics[width=80mm]{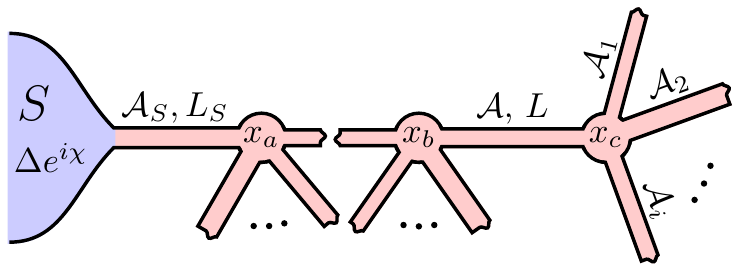}
\caption{An example of the network formed by quasi-one-dimensional normal wires attached to a superconducting terminal.}
\label{F-fig}
\end{figure}

Let us assume that electron energy $\varepsilon$ exceeds the Thouless energy for any wire segment. In this case the Usadel equation can be integrated straightforwardly from the superconducting terminal deep into normal wire network. At relatively short distances $x \ll L_S$ away from a superconducting terminal one can safely disregard the effect of all other terminals and the anomalous Green function can be expressed in the form
\begin{gather}
F^R = -i \dfrac{4 y (1- y^2)}{(1+y^2)^2} e^{i\chi},
\quad
y = a_{S} e^{-\sqrt{-2i\varepsilon/D}x},
\\
a_{S} (\varepsilon) = 
\tan\left[\dfrac{1}{4}\arcsin \dfrac{|\Delta|}{\sqrt{|\Delta|^2 - \varepsilon^2}}\right],
\label{FRSN}
\end{gather}
where $\Delta = |\Delta|e^{i\chi}$ is the order parameter in the corresponding superconducting terminal.

At distances away from $S$-terminals exceeding $\sqrt{D/|\varepsilon|}$ one can linearize the Usadel equation everywhere inside the network of normal wires 
\begin{equation}
(F^R)''  - k^2 F^R=0, 
\quad
k = \sqrt{-2i\varepsilon/D}
\label{FRlinear}
\end{equation} 
and the anomalous Green functions take exponentially small values 
\begin{equation}
F^R = -4 i a_{S} e^{-\sqrt{-2i\varepsilon/D}x} e^{i\chi}.
\label{FRSN2}
\end{equation}
At larger distances from the superconducting terminal the anomalous Green function $F$ obeys Eq. \eqref{FRlinear} which can be integrated step by step along the normal wire network. 

As an illustration, let us consider an arbitrary normal wire segment $x_bx_c$ with length $L$ and cross section area $\mathcal{A}$ (see Fig. \ref{F-fig}). A general solution of Eq. \eqref{FRlinear} reads
\begin{equation}
F^R = C_- e^{-k(x-x_b)} + C_+e^{k(x-x_b)},
\label{FC}
\end{equation}
where one obviously has $C_+ \ll C_-$. Hence, with a good accuracy we can set 
\begin{equation}
C_- = F^R(x_b).
\label{C-}
\end{equation}
The prefactor $C_+$ can be derived from the boundary conditions at the point $x=x_c$ which include the continuity condition for the Green function $F^R$ combined with the equation
\begin{equation}
\mathcal{A} (F^R(x_c))' = \sum_i \mathcal{A}_i (F^R_i(x_c))',
\end{equation}
where the index $i$ enumerates all normal wires attached to the wire segment $x_bx_c$ at the node $x_c$. As a result, we obtain
\begin{equation}
C_+ = F^R(x_b) \dfrac{\mathcal{A} - \sum_i \mathcal{A}_i}{\mathcal{A} + \sum_i \mathcal{A}_i} e^{-2kL}
\label{C+}
\end{equation}
and, hence,
\begin{equation}
F^R(x_c) = F^R(x_b) \dfrac{2 \mathcal{A}}{\mathcal{A} + \sum_i \mathcal{A}_i}e^{-kL}.
\end{equation}
This equation together with Eqs. \eqref{FC}, \eqref{C-}, \eqref{C+} allows one to evaluate the anomalous Green functions everywhere inside our structure. 

Obviously, the above analysis can easily be generalized to the network of normal wires connected to several superconducting terminals.
In this case the resulting anomalous Green function everywhere inside this network is given by a simple superposition of the contributions from each of these terminals.

Turning back to the structure depicted in Fig. \ref{snns3-fig} and making use of the above results, in the high energy limit $ |\varepsilon| \gg E_{\mathrm{Th}}$ we get
\begin{multline}
F^R =
-\dfrac{\kappa \mathcal{A}_S e^{-kL_S} e^{i\chi_1}}{
\mathcal{A}_S + \mathcal{A}_N + \mathcal{A}_p
}
e^{-kx}
\\-
\dfrac{\kappa \mathcal{A}_S e^{-kL_S} e^{i\chi_2}}{
\mathcal{A}_S + \mathcal{A}_N + \mathcal{A}_p
}
e^{-k(L_p-x)}
,
\label{Fp}
\end{multline}
in the central wire and
\begin{multline}
F^R =
-\dfrac{\kappa \mathcal{A}_S e^{-kL_S} e^{i\chi_1}}{
\mathcal{A}_S + \mathcal{A}_N + \mathcal{A}_p
}
e^{-kx}
\\-
\dfrac{\kappa\mathcal{A}_S  e^{-kL_S} e^{i\chi_2}}{
\mathcal{A}_S + \mathcal{A}_N + \mathcal{A}_p
}
e^{-k L_p}
\dfrac{2 \mathcal{A}_p}{\mathcal{A}_S + \mathcal{A}_N + \mathcal{A}_p}
e^{-kx}
,
\label{FN}
\end{multline}
in the wire connected to the normal terminal $N_1$. Here $x$ is a coordinate along the wire from the crossing point $p_1$,
and $\kappa = 8i/(1+\sqrt{2})$. The anomalous Green function $\tilde F^R$ is recovered from the expressions \eqref{Fp} and \eqref{FN} by inverting their overall sign and performing the phase inversion $\chi_{1,2}\leftrightarrow -\chi_{1,2}$. Substituting these results for $F^R$ and $\tilde F^R$ into Eq. \eqref{jE} and evaluating the integral in Eq. \eqref{IJ} we arrive at Eq. \eqref{IJ2} for the equilibrium Josephson current in our structure.

The spectral conductances $G_j$ and $G_N^{\mathcal{Y}}$ can be evaluated analogously. In the high energy limit we obtain 
\begin{multline}
G_j
=
\frac{128}{3+2\sqrt{2}}\dfrac{\mathcal{A}_S^2 \mathcal{A}_p \sigma}{
(\mathcal{A}_S + \mathcal{A}_N + \mathcal{A}_p)^2}
\sin \chi
\\\times
k'e^{-k' L}
\left(
\cos k' L  - \sin k' L 
\right)\sgn \varepsilon
\label{Gjapprox}
\end{multline}
and  
\begin{multline}
G_N^{\mathcal{Y}} \simeq
-
G_N^n
\dfrac{64}{3+2\sqrt{2}}\dfrac{\mathcal{A}_S^2 \mathcal{A}_p}{
(\mathcal{A}_S + \mathcal{A}_N + \mathcal{A}_p)^3}
\\\times
\sin \chi
e^{-k' L}
\sin k' L_p  
\dfrac{1}{k' L_N}
\sgn \varepsilon.
\label{GMasym}
\end{multline}
where we define $L = 2L_S + L_p$ and $k' = \sqrt{|\varepsilon|/D}$.

Finally, we specify the expression for $\tilde G$ which enters Eq. \eqref{ISlT}. It reads
\begin{widetext}
\begin{multline}
\tilde G 
=
E_{\mathrm{Th}}^2
\Biggl[
\dfrac{1}{4}
\dfrac{ (G_N^{\mathcal{Y}})' G_j' G_S^n 2 G_p^L(0) + (G_j')^2 G_N^n G_N^L(0)
}{
G_S^n G_N^n [2 G_p^L(0) + G_N^L(0)]
}
\\+
\dfrac{[G_S'' G_p^n - G_S^n (G_p^T)'']  G_N^n G_N^L(0) 
- 
(G_S^n + 2 G_p^n)^2 G_j' (G_N^{\mathcal{Y}})'
}{G_N^L(0) (G_N^n + G_S^n + 2 G_p^n)(G_S^n + 2 G_p^n)}
\Biggr],
\label{ISshort}
\end{multline}
\end{widetext}
where $G_S''$ and $(G_p^T)''$ represent the second derivatives of the corresponding spectral conductances with respect to energy at $\varepsilon =0$.

\section{Asymmetric interferometers}
\label{Aasym}
Let us apply our circuit theory approach to the analysis of subgap electron transport in Andreev interferometers displayed in Fig. \ref{snns3-fig} making no assumption about their symmetry. Resolving the system of matrix equations \eqref{IS1}-\eqref{Ip2} we can derive the expressions for the matrix currents as functions of the electron distribution functions inside the terminals. We obtain  
\begin{widetext}
\begin{gather}
\hat I_{N_1}
=
\hat G_{N_1} 
\hat P
\Biggl[
\hat G_{S_1} + \hat G_{S_2}
+
( \hat G_{S_2} + \hat G_{N_2} - \hat \tau_+ G_j)
(\hat G_p + \hat \tau_1 G_j /2)^{-1}
( \hat G_{S_1} - \hat \tau_- G_j)
\Biggr]
\hat H_{N_1}
+
\hat A ( \hat H_{N_1} - \hat H_{N_2} ),
\label{IN1asymmatr}
\\
\hat I_{N_2}
=
\hat G_{N_2} 
\hat \tau_3
\hat P^T
\hat \tau_3
\Biggl[
\hat G_{S_1} + \hat G_{S_2}
+
( \hat G_{S_1} + \hat G_{N_1} + \hat \tau_+ G_j)
(\hat G_p - \hat \tau_1 G_j /2)^{-1}
( \hat G_{S_2} + \hat \tau_- G_j)
\Biggr]
\hat H_{N_2}
-
\hat \tau_3
\hat A^T
\hat \tau_3
( \hat H_{N_1} - \hat H_{N_2} ),
\label{IN2asymmatr}
\end{gather}
where the matrices $\hat P$ and $\hat A$ read
\begin{gather}
\hat P =
\Biggl[
\hat G_{S_1} + \hat G_{S_2}+ \hat G_{N_1} + \hat G_{N_2}
+
( \hat G_{S_2} + \hat G_{N_2} - \hat \tau_+ G_j)
(\hat G_p + \hat \tau_1 G_j /2)^{-1}
( \hat G_{S_1} + \hat G_{N_1} - \hat \tau_- G_j)
\Biggr]^{-1},
\quad
\hat A = \hat G_{N_1}  \hat P \hat G_{N_2}.
\end{gather}
The currents $I_{N_{1,2}}$ flowing in the normal wires connected to the terminals $N_1$ and $N_2$ are defined by the formula
\begin{equation}
I_{N_{1,2}}
=
\dfrac{1}{2}
\int I^T_{1,2} d \varepsilon.
\end{equation}
Obviously, both these currents vanish provided the terminals are disconnected from any external circuit, i.e. the conditions \eqref{zcc} apply
which determine the thermoelectric voltages $V_1$ and $V_2$ induced at these terminals. In the high temperature limit $T_{1,2} \gg E_{\mathrm{Th}}$ the calculation of $V_1$ and $V_2$ is simplified since the prefactors in front of the distribution functions $h^T_{N_{1,2}}$ entering the general expressions for $I_{N_{1,2}}$ can be replaced by their normal state values. Then we get
\begin{multline}
0 = I_{N_1}
=
G_{N_1}^n
\dfrac{G_p^n( G_{S_1}^n + G_{S_2}^n )
+
( G_{S_2}^n + G_{N_2}^n) G_{S_1}^n }{(G_{S_1}^n + G_{S_2}^n + G_{N_1}^n + G_{N_2}^n)G_p^n
+
( G_{S_2}^n + G_{N_2}^n) ( G_{S_1}^n + G_{N_1}^n)}
V_1
\\+
G_{N_1}^n G_{N_2}^n
\dfrac{G_p^n }{
[G_p^n (G_{S_1}^n + G_{S_2}^n + G_{N_1}^n + G_{N_2}^n) + (G_{S_1}^n + G_{N_1}^n) (G_{S_2}^n + G_{N_2}^n)]}
(V_1 - V_2)
+
\dfrac{1}{2e}
\int 
A_{12} [h^L_{N_1} - h^L_{N_2}] d \varepsilon,
\label{V1}
\end{multline}
and
\begin{multline}
0=I_{N_2}
=
G_{N_2}^n
\dfrac{G_p^n( G_{S_1}^n + G_{S_2}^n )
+
( G_{S_1}^n + G_{N_1}^n) G_{S_2}^n }{(G_{S_1}^n + G_{S_2}^n + G_{N_1}^n + G_{N_2}^n)G_p^n
+
( G_{S_2}^n + G_{N_2}^n) ( G_{S_1}^n + G_{N_1}^n)}
V_2
\\-
G_{N_1}^n G_{N_2}^n
\dfrac{G_p^n }{
[G_p^n (G_{S_1}^n + G_{S_2}^n + G_{N_1}^n + G_{N_2}^n) + (G_{S_1}^n + G_{N_1}^n) (G_{S_2}^n + G_{N_2}^n)]}
(V_1 - V_2)
+
\dfrac{1}{2e}
\int 
A_{21}
[h^L_{N_1} - h^L_{N_2}] d \varepsilon,
\label{V2}
\end{multline}
where $A_{12}$ and $A_{21}$ denote the off-diagonal elements of the matrix $\hat A$
\begin{equation}
\hat A =
\begin{pmatrix}
A_{11} & A_{12}
\\
A_{21} & A_{22}
\end{pmatrix}.
\end{equation}
Resolving Eqs. \eqref{V1} and \eqref{V2}, we obtain
\begin{gather}
V_1
= -
\dfrac{1}{2e}
\int 
\dfrac{[G_p^n( G_{S_1}^n + G_{S_2}^n + G_{N_1}^n)
+
( G_{S_1}^n + G_{N_1}^n) G_{S_2}^n]
A_{12}
+
G_p^n G_{N_1}^n 
A_{21}
}{G_{N_1}^n
[(G_{S_1}^n + G_{S_2}^n )G_p^n + G_{S_2}^n G_{S_1}^n ]}
[h^L_{N_1} - h^L_{N_2}] d \varepsilon ,
\label{V1-2}
\\
V_2
= -
\dfrac{1}{2e}
\int 
\dfrac{[G_p^n( G_{S_1}^n + G_{S_2}^n + G_{N_2}^n)
+
( G_{S_2}^n + G_{N_2}^n) G_{S_1}^n]
A_{21}
+
G_p^n G_{N_2}^n 
A_{12}
}{G_{N_2}^n
[(G_{S_1}^n + G_{S_2}^n )G_p^n + G_{S_2}^n G_{S_1}^n ]}
[h^L_{N_1} - h^L_{N_2}] d \varepsilon ,
\label{V2-2}
\end{gather}
Here with a good accuracy one can neglect the voltage dependence of the distribution functions $h^L_{N_{1,2}}$ replacing them by their equilibrium values $h^L_{N_{1,2}} = \tanh [\varepsilon /(2T_{1,2})]$. It is convenient to identically rewrite the matrix elements $A_{12}$ and $A_{21}$ in the form
\begin{gather}
A_{12} = \dfrac{1}{Q_n}(G_{N_1}^n)^2  (G_{S_2}^n + G_{N_2}^n + 2 G_p^n) G_{N_2}^n  G_j/2
+
\Biggl[
A_{12} - \dfrac{1}{Q_n}(G_{N_1}^n)^2  (G_{S_2}^n + G_{N_2}^n + 2 G_p^n) G_{N_2}^n  G_j/2
\Biggr],
\\
A_{21} = \dfrac{1}{Q_n}(G_{N_2}^n)^2  (G_{S_1}^n + G_{N_1}^n + 2 G_p^n) G_{N_1}^n  G_j/2
+
\Biggl[
A_{21} - \dfrac{1}{Q_n}(G_{N_2}^n)^2  (G_{S_1}^n + G_{N_1}^n + 2 G_p^n) G_{N_1}^n  G_j/2
\Biggr],
\end{gather}
where we explicitly extracted $G_j$-terms with the prefactors replaced by their normal state values. Here  
\begin{equation}
Q_n =
[G_p^n (G_{S_1}^n + G_{S_2}^n + G_{N_1}^n + G_{N_2}^n) + (G_{S_1}^n + G_{N_1}^n) (G_{S_2}^n + G_{N_2}^n)]
[G_p^n (G_{N_1}^n + G_{N_2}^n) + G_{N_1}^n G_{N_2}^n]
\end{equation}
is the normal state value of the function 
\begin{gather}
Q = \det \begin{vmatrix}
\hat G_{S_1} + \hat G_{S_2}+ \hat G_{N_1} + \hat G_{N_2} & \hat G_{S_2} + \hat G_{N_2} - \hat \tau_+ G_j \\
\hat G_{S_1} + \hat G_{N_1} - \hat \tau_- G_j & -(\hat G_p + \hat \tau_1 G_j /2)
\end{vmatrix}.
\end{gather}
With this in mind Eqs. \eqref{V1-2} and \eqref{V2-2} can be rewritten as
\begin{multline}
V_1
=
\dfrac{G_{N_1}^n G_{N_2}^n}{2}
\dfrac{2 G_p^n + G_{S_2}^n}{
[G_p^n (G_{N_1}^n + G_{N_2}^n ) + G_{N_1}^n G_{N_2}^n ]
[G_p^n (G_{S_1}^n + G_{S_2}^n ) + G_{S_2}^n G_{S_1}^n ]}
[I_J(T_1, \chi) - I_J(T_2, \chi)] 
\\-
\dfrac{1}{4e}
\left(
\dfrac{1}{T_1} - \dfrac{1}{T_2}
\right)
\dfrac{1}{G_{N_1}^n
[(G_{S_1}^n + G_{S_2}^n )G_p^n + G_{S_2}^n G_{S_1}^n ]}
\int 
\Biggl\{
[G_p^n( G_{S_1}^n + G_{S_2}^n + G_{N_1}^n)
+
( G_{S_1}^n + G_{N_1}^n) G_{S_2}^n]
A_{12}
+
G_p^n G_{N_1}^n 
A_{21}
\\-
\dfrac{G_{N_1}^n G_{N_2}^n}{2}
\dfrac{ G_{N_1}^n (2 G_p^n + G_{S_2}^n)}{[G_p^n (G_{N_1}^n + G_{N_2}^n) + G_{N_1}^n G_{N_2}^n]}G_j
\Biggr\}
\varepsilon d \varepsilon,
\label{V1-3}
\end{multline}
\begin{multline}
V_2
=
\dfrac{G_{N_1}^n G_{N_2}^n}{2}
\dfrac{2 G_p^n + G_{S_1}^n}{
[G_p^n (G_{N_1}^n + G_{N_2}^n ) + G_{N_1}^n G_{N_2}^n ]
[G_p^n (G_{S_1}^n + G_{S_2}^n ) + G_{S_2}^n G_{S_1}^n ]}
[I_J(T_1, \chi) - I_J(T_2, \chi)] 
\\-
\dfrac{1}{4e}
\left(
\dfrac{1}{T_1} - \dfrac{1}{T_2}
\right)
\dfrac{1}{G_{N_2}^n
[(G_{S_1}^n + G_{S_2}^n )G_p^n + G_{S_2}^n G_{S_1}^n ]}
\int 
\Biggl\{
[G_p^n( G_{S_1}^n + G_{S_2}^n + G_{N_2}^n)
+
( G_{S_2}^n + G_{N_2}^n) G_{S_1}^n]
A_{21}
+
G_p^n G_{N_2}^n 
A_{12}
\\-
\dfrac{G_{N_1}^n G_{N_2}^n}{2}
\dfrac{ G_{N_2}^n (2 G_p^n + G_{S_1}^n)}{[G_p^n (G_{N_1}^n + G_{N_2}^n) + G_{N_1}^n G_{N_2}^n]}G_j
\Biggr\}
\varepsilon d \varepsilon.
\label{V2-3}
\end{multline}
\end{widetext}

Quasi-equilibrium contributions containing the difference between the Josephson currents $I_J(T_1, \chi) - I_J(T_2, \chi)$  in Eqs. \eqref{V1-3} and  \eqref{V2-3} coincide with the corresponding terms derived in Ref. \cite{VH}.

The supercurrent $I_S$ flowing in our circuit can be derived with the aid of the formula
\begin{equation}
I_S = \dfrac{1}{2}
\int I_{S_1}^T d \varepsilon.
\end{equation}
Here $I_{S_1}^T$ is the corresponding component of the matrix current $\hat I_{S_1}$ obtained from Eqs. \eqref{IS1}-\eqref{Ip2} in the form
\begin{multline}
e \hat I_{S_1}=
- (\hat G_{S_1} + \hat \tau_+ G_j) \hat H_{N_1}
+
(\hat G_{S_1} + \hat \tau_+ G_j)
\hat P
\Bigl[
\hat G_{S_1} + \hat G_{S_2}
\\+
( \hat G_{S_2} + \hat G_{N_2} - \hat \tau_+ G_j)
(\hat G_p + \hat \tau_1 G_j /2)^{-1}
( \hat G_{S_1} - \hat \tau_- G_j)
\Bigr]
\hat H_{N_1}
\\+
(\hat G_{S_1} + \hat \tau_+ G_j)
\hat P \hat G_{N_2} ( \hat H_{N_1} - \hat H_{N_2}).
\label{IS1asymmatr}
\end{multline}

At high temperatures $T_{1,2} \gg E_{\mathrm{Th}}$ the supercurrent can be evaluated explicitly in exactly the same manner as the thermoelectric voltages $V_{1,2}$. In this limit we obtain
\begin{widetext}
\begin{multline}
I_S= 
\dfrac{
(G_{S_1}^n + G_{S_2}^n ) G_{N_1}^n (G_p^n)^2  
+ 
[G_{N_1}^n G_{N_2}^n G_{S_2}^n + G_{S_1}^n G_{S_2}^n G_{N_1}^n  ] G_p^n 
+ 
\dfrac{1}{2} G_{S_1}^n G_{S_2}^n G_{N_1}^n G_{N_2}^n }{
[G_p^n (G_{N_1}^n + G_{N_2}^n) + G_{N_1}^n G_{N_2}^n] [(G_{S_1}^n + G_{S_2}^n )G_p^n + G_{S_2}^n G_{S_1}^n ]}
I_J(T_1, \chi)
\\+
\dfrac{
(G_{S_1}^n + G_{S_2}^n ) G_{N_2}^n (G_p^n)^2  
+ 
[G_{N_1}^n G_{N_2}^n G_{S_1}^n + G_{S_1}^n G_{S_2}^n G_{N_2}^n  ] G_p^n 
+ 
\dfrac{1}{2} G_{S_1}^n G_{S_2}^n G_{N_1}^n G_{N_2}^n }{
[G_p^n (G_{N_1}^n + G_{N_2}^n) + G_{N_1}^n G_{N_2}^n] [(G_{S_1}^n + G_{S_2}^n )G_p^n + G_{S_2}^n G_{S_1}^n ]}
I_J(T_2, \chi)
\\+
\dfrac{1}{4e}
\left(
\dfrac{1}{T_1}
-
\dfrac{1}{T_2}
\right)
\int
\Biggl\{
\dfrac{
(G_{S_1}^n + G_{S_2}^n ) G_{N_1}^n (G_p^n)^2  
+ 
[G_{N_1}^n G_{N_2}^n G_{S_2}^n + G_{S_1}^n G_{S_2}^n G_{N_1}^n  ] G_p^n 
+ 
\dfrac{1}{2} G_{S_1}^n G_{S_2}^n G_{N_1}^n G_{N_2}^n }{
[G_p^n (G_{N_1}^n + G_{N_2}^n) + G_{N_1}^n G_{N_2}^n] [(G_{S_1}^n + G_{S_2}^n )G_p^n + G_{S_2}^n G_{S_1}^n ]}
\\\times
\Biggl[
G_{S_1}^n
\dfrac{
(G_p^n + G_{S_2}^n) A_{12}
+
G_p^n A_{21}
}{(G_{S_1}^n + G_{S_2}^n )G_p^n + G_{S_2}^n G_{S_1}^n  }
+
P_{11} G_{S_1} G_{N_2}^{\mathcal{Y}} + P_{12} G_{S_1} G_{N_2}^L + P_{21} G_j G_{N_2}^{\mathcal{Y}} + P_{22} G_j G_{N_2}^L
\Biggr]
\\+
\dfrac{
(G_{S_1}^n + G_{S_2}^n ) G_{N_2}^n (G_p^n)^2  
+ 
[G_{N_1}^n G_{N_2}^n G_{S_1}^n + G_{S_1}^n G_{S_2}^n G_{N_2}^n  ] G_p^n 
+ 
\dfrac{1}{2} G_{S_1}^n G_{S_2}^n G_{N_1}^n G_{N_2}^n }{
[G_p^n (G_{N_1}^n + G_{N_2}^n) + G_{N_1}^n G_{N_2}^n] [(G_{S_1}^n + G_{S_2}^n )G_p^n + G_{S_2}^n G_{S_1}^n ]}
\\\times
\Biggl[P_{11} G_{S_2} G_{N_1}^{\mathcal{Y}} - P_{21} G_{S_2} G_{N_1}^L + P_{12} G_j G_{N_1}^{\mathcal{Y}} - P_{22} G_j G_{N_1}^L -
G_{S_2}^n
\dfrac{G_p^n A_{12} + (G_p^n + G_{S_1}) A_{21} }{
(G_{S_1}^n + G_{S_2}^n )G_p^n + G_{S_2}^n G_{S_1}^n }\Biggr]
\Biggr\}
\varepsilon
d\varepsilon.
\label{ISasym}
\end{multline}
\end{widetext}

For completeness, let us also consider the low temperature limit $T_{1,2} \ll E_{\mathrm{Th}}$. In order to evaluate thermoelectric voltages
$V_1$ and $V_2$ in this limit it suffices to replace $G^T$ and $G^L$ components of the matrix conductances by their zero energy values and set
\begin{equation}
G_j \approx G_j' \varepsilon,
\quad
G_X^{\mathcal{Y}} \approx (G_X^{\mathcal{Y}})' \varepsilon,
\quad X=N_{1,2},p.
\end{equation}
With the same accuracy we can neglect products of the $G_j$ and $G^{\mathcal{Y}}$ conductances since they have higher power of energy. At zero energy $G^T$ components of the matrix conductances are known to exactly coincide with corresponding normal state conductances \cite{BWBSZ1999,Nazarov96,golubov1997coherent}. With this in mind one can demonstrate that Eqs. \eqref{V1-2} and \eqref{V2-2} remain applicable also in the low temperature limit. Evaluating $A_{12}$ and $A_{21}$, we get
\begin{widetext}
\begin{gather}
V_1
=
\dfrac{\pi^2}{12}
\dfrac{G_{N_2}^L 
\left\{ 2 [ (G_{S_1}^n + G_{S_2}^n) G_p^n  + G_{S_1}^n G_{S_2}^n ] G_p^L  (G_{N_1}^{\mathcal{Y}})'
+
G_{N_1}^n (2G_p^n + G_{S_2}^n) G_{N_1}^L G_j'
+
2 G_{N_1}^n G_{S_2}^n G_{N_1}^L (G_p^{\mathcal{Y}})'
\right\}}{e G_{N_1}^n
[(G_{S_1}^n + G_{S_2}^n )G_p^n + G_{S_2}^n G_{S_1}^n ][G_{N_1}^L + G_{N_2}^L) G_p^L + G_{N_1}^L G_{N_2}^L ]}
(T_1^2 - T_2^2),
\label{V1theorshort}
\\
V_2
=
\dfrac{\pi^2}{12}
\dfrac{G_{N_1}^L 
\left\{ - 2 [ (G_{S_1}^n + G_{S_2}^n) G_p^n  + G_{S_1}^n G_{S_2}^n ] G_p^L  (G_{N_2}^{\mathcal{Y}})'
+
G_{N_2}^n (2G_p^n + G_{S_1}^n) G_{N_2}^L (G_j)'
-
2 G_{N_2}^n G_{S_1}^n G_{N_2}^L (G_p^{\mathcal{Y}})'
\right\}}{e G_{N_2}^n
[(G_{S_1}^n + G_{S_2}^n )G_p^n + G_{S_2}^n G_{S_1}^n ][G_{N_1}^L + G_{N_2}^L) G_p^L + G_{N_1}^L G_{N_2}^L ]}
(T_1^2 - T_2^2),
\label{V2theorshort}
\end{gather}
where the spectral conductances $G_{N_1}^L$ $G_{N_2}^L$ and $G_p^L$ are evaluated at $\varepsilon=0$.

The Josephson current $I_S$ is evaluated in exactly the same way. As a result, we obtain
\begin{multline}
I_S =
\Biggl[
\dfrac{ 
(G_{S_1}^n + G_{S_2}^n ) G_{N_1}^L G_p^L G_p^n 
+ 
G_{S_2}^n G_{N_1}^L G_{N_2}^L G_p^n 
+
G_{S_2}^n G_{S_1}^n  G_{N_1}^L G_p^L
+ 
G_{S_1}^n G_{S_2}^n G_{N_1}^L G_{N_2}^L /2 
}{
[(G_{S_1}^n + G_{S_2}^n )G_p^n + G_{S_2}^n G_{S_1}^n]
[(G_{N_1}^L + G_{N_2}^L) G_p^L + G_{N_1}^L G_{N_2}^L]
}
\Biggr]
I_J(\chi, T_1)
\\+
\Biggl[
\dfrac{ 
(G_{S_1}^n + G_{S_2}^n ) G_{N_2}^L G_p^L G_p^n 
+ 
G_{S_1}^n G_{N_1}^L G_{N_2}^L G_p^n 
+
G_{S_2}^n G_{S_1}^n  G_{N_2}^L G_p^L
+ 
G_{S_1}^n G_{S_2}^n G_{N_1}^L G_{N_2}^L /2 
}{
[(G_{S_1}^n + G_{S_2}^n )G_p^n + G_{S_2}^n G_{S_1}^n]
[(G_{N_1}^L + G_{N_2}^L) G_p^L + G_{N_1}^L G_{N_2}^L]
}
\Biggr]
I_J(\chi, T_2)
\\-
\dfrac{\pi^2}{6}\dfrac{G_{S_1}^n G_{S_2}^n}{
(G_{S_1}^n + G_{S_2}^n )G_p^n + G_{S_2}^n G_{S_1}^n }
\dfrac{G_{N_1}^L G_{N_2}^L}{(G_{N_1}^L + G_{N_2}^L) G_p^L + G_{N_1}^L G_{N_2}^L} (G_p^{\mathcal{Y}})'
(T_1^2 - T_2^2).
\label{Jas}
\end{multline}
\end{widetext}

\end{document}